\documentclass[12pt]{article}
\usepackage[mathscr]{eucal}
\usepackage{epsfig,amsfonts}
\usepackage{amsmath}
\usepackage{amsthm,amssymb}
\usepackage{graphicx}
\usepackage{hhline}
\usepackage{relsize}
\usepackage{cite}
\usepackage{psfrag}
\usepackage{mathrsfs} 
\usepackage{bm}
\usepackage{array}
\usepackage{tikz}
\makeatletter
\@addtoreset{equation}{section}
\makeatother

\def\bea{\begin{eqnarray}}
\def\eea{\end{eqnarray}}
\def\be{\begin{equation}}
\def\ee{\end{equation}}

\def\be{\begin{equation}}
\def\ee{\end{equation}}
\def\bdm{\begin{displaymath}}
\def\edm{\end{displaymath}}
\def\bea{\begin{eqnarray}}
\def\eea{\end{eqnarray}}

\def\ri{{\rm i}}

\def\XXint#1#2#3{{\setbox0=\hbox{$#1{#2#3}{\int}$}
    \vcenter{\hbox{$#2#3$}}\kern-.5\wd0}}

\newcommand{\rd}{\mbox{d}}
\newcommand{\re}{\mbox{e}}

\DeclareMathAlphabet{\mathpzc}{OT1}{pzc}{m}{it}
\usepackage{hyperref}

\begin{document}

\begin{titlepage}
\begin{flushright}
$\phantom{{\it tresrtfdsgqw }}$\\
\end{flushright}

\vspace{0.8cm}

\begin{center}
\begin{LARGE}

{\bf  On the Yang-Baxter Poisson algebra in
\\[0.5cm] 
non-ultralocal integrable systems
}
\end{LARGE}
\vspace{1.3cm}
\begin{large}

{\bf Vladimir V. Bazhanov$^{1}$,  Gleb A.  Kotousov$^{1,2}$  \\
\bigskip
and Sergei  L. Lukyanov$^{2,3}$}

\end{large}

\vspace{1.cm}
$^1$Department of Theoretical Physics\\
         Research School of Physics and Engineering\\
    Australian National University, Canberra, ACT 2601, Australia\\\ \\
${}^{2}$NHETC, Department of Physics and Astronomy\\
     Rutgers University\\
     Piscataway, NJ 08855-0849, USA\\
\vspace{.2cm}
and\\
\vspace{.2cm}
${}^{3}$Kharkevich Institute for Information Transmission Problems\\
Moscow, 127994, Russia
\vspace{1.0cm}

\end{center}

\begin{center}
\centerline{\bf Abstract} \vspace{.8cm}

\parbox{13cm}{%
A common approach to the quantization of integrable models 
starts with the formal substitution of the Yang-Baxter Poisson algebra with 
its quantum version.
However it is
 difficult to discern 
 the  presence of such an algebra
 for the so-called non-ultralocal models. 
 The latter includes the class of
 non-linear sigma models which are most interesting
 from the point of view of applications.
In this work, we investigate the emergence of the Yang-Baxter Poisson algebra
in a non-ultralocal system
which is related to integrable deformations of the Principal Chiral Field.
}
\end{center}
\vspace{.8cm}

\vfill

\end{titlepage}
\setcounter{page}{2}

\section{Introduction}
\begin{figure}[t]
\centering
\scalebox{0.7}{
\begin{tikzpicture}
\draw[line width = 0.5mm]  (0,0) -- (3.5,7);
\draw[line width = 0.5mm]  (-0.5,0.346) -- (5,4.154);
\draw[line width = 0.5mm] (0,6) -- (2,0);
\draw[line width = 0.5mm] (0.7,6.2) -- (2.7,0.2);
\draw[line width = 0.5mm] (1.4,6.4) -- (3.4,0.4);
\draw[line width = 0.5mm] (2.1,6.6) -- (4.1,0.6);
\draw[line width = 0.5mm] (2.8,6.8) -- (4.8,0.8);
\draw[line width = 0.8mm] (8.3,3.8) -- (9.1,3.8);
\draw[line width = 0.8mm] (8.3,3.4) -- (9.1,3.4);
\draw[line width = 0.5mm]  (13.6,0) -- (17.1,7);
\draw[line width = 0.5mm]  (12.1,2.846) -- (17.6,6.654);
\draw[line width = 0.5mm] (12.3,6) -- (14.3,0);
\draw[line width = 0.5mm] (13,6.2) -- (15,0.2);
\draw[line width = 0.5mm] (13.7,6.4) -- (15.7,0.4);
\draw[line width = 0.5mm] (14.4,6.6) -- (16.4,0.6);
\draw[line width = 0.5mm] (15.1,6.8) -- (17.1,0.8);
\draw[blue,fill=blue] (0.529,1.059) circle (.7ex);
\draw[blue,fill=blue] (1.4375,1.69) circle (.7ex);
\draw[blue,fill=blue] (2.06,2.12) circle (.7ex);
\draw[blue,fill=blue] (2.68,2.55) circle (.7ex);
\draw[blue,fill=blue] (3.306,2.98) circle (.7ex);
\draw[blue,fill=blue] (3.93,3.4125) circle (.7ex);
\draw[blue,fill=blue] (1.66,3.32) circle (.7ex);
\draw[blue,fill=blue] (2.12,4.24) circle (.7ex);
\draw[blue,fill=blue] (2.58,5.16) circle (.7ex);
\draw[blue,fill=blue] (3.04,6.08) circle (.7ex);
\draw[blue,fill=blue] (1.2,2.4) circle (.7ex);
\draw[blue,fill=blue] (16.57,5.94) circle (.7ex);
\draw[blue,fill=blue] (14.02,0.84) circle (.7ex);
\draw[blue,fill=blue] (14.48,1.76) circle (.7ex);
\draw[blue,fill=blue] (14.94,2.68) circle (.7ex);
\draw[blue,fill=blue] (15.4,3.6) circle (.7ex);
\draw[blue,fill=blue] (15.86,4.52) circle (.7ex);
\draw[blue,fill=blue] (13.117,3.55) circle (.7ex);
\draw[blue,fill=blue] (13.74,3.98) circle (.7ex);
\draw[blue,fill=blue] (14.36,4.4125) circle (.7ex);
\draw[blue,fill=blue] (14.9854,4.84375) circle (.7ex);
\draw[blue,fill=blue] (15.6083,5.275) circle (.7ex);
%
%
\node at (4,6.5) {\Large $1$};
\node at (5.5,3.7) {\Large $2$};
\node at (13,0.3) {\Large $1$};
\node at (11.7,3.2) {\Large $2$};
\end{tikzpicture}
}
\caption{
A graphical representation of the Yang-Baxter relation in solvable lattice models.
\label{fig1}}
\end{figure}
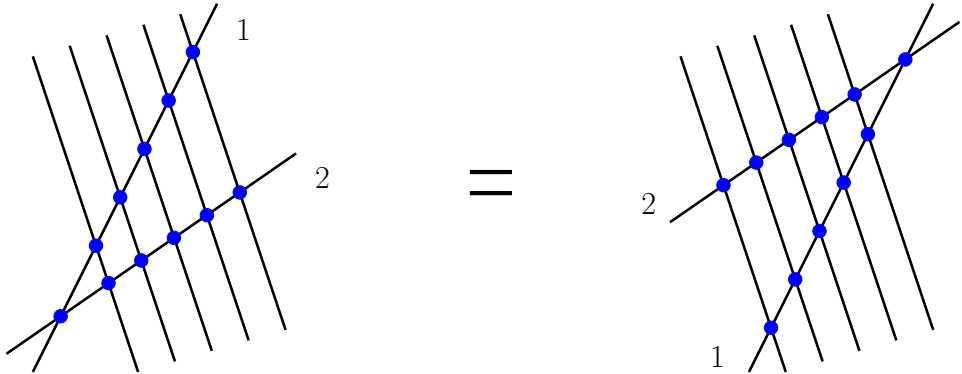
\noindent
Throughout the development of integrability, there has been a fruitful exchange of ideas 
and methods centered around the mathematical structure commonly known as the Yang-Baxter algebra
\be\label{RTT1}
{\boldsymbol R}(\lambda_2/\lambda_1)\,\big({\boldsymbol M}(\lambda_1)\otimes{\boldsymbol 1}\big)\,
\big({\boldsymbol 1}\otimes{\boldsymbol M}(\lambda_2)\big)=
\big({\boldsymbol 1}\otimes{\boldsymbol M}(\lambda_2)\big)\,
\big({\boldsymbol M}(\lambda_1)\otimes{\boldsymbol 1}\big)
{\boldsymbol R}(\lambda_2/\lambda_1)\ .
\ee
It appeared in the context of
lattice systems \cite{Baxter:1982zz} with ${\boldsymbol M}$ being a matrix built from the
local statistical weights which satisfy a local Yang-Baxter equation (see fig.\,\ref{fig1}).
The fundamental r$\hat{{\rm o}}$le of the Yang-Baxter algebra in the context of $1+1$ dimensional 
classically integrable field theory was first pointed out by Sklyanin \cite{Sklyanin:1979gh} and further developed
in the works of the Leningrad school \cite{Faddeev:1987ph}.
It was observed that for many partial differential equations admitting the zero curvature representation,
the canonical Poisson structure yields the
equal-time Poisson brackets
\be\label{PB0}
\big\{{\boldsymbol {A}}_x(x|\lambda_1) \begin{array}{ccc} \\[-0.4cm] \otimes \\[-0.35cm] , \end{array}  {\boldsymbol { A}}_x (y|\lambda_2)\big\}
= \big[
{\boldsymbol { A}}_x(x|\lambda_1)\otimes {\boldsymbol  1}+
 {\boldsymbol 1}\otimes {\boldsymbol {A}}_x(y|\lambda_2), {\boldsymbol r}(\lambda_1/\lambda_2)\big]\, \delta(x-y)
\ee
for the $x$-component of the flat connection.
The ``ultralocal'' relations \eqref{PB0} imply that the  monodromy matrix, 
\be\label{mod2}
{\boldsymbol M}(\lambda)=
\overset{\leftarrow}{{\cal P}}\exp\bigg(\int_{0}^{R}{\rm d} x\ 
 {\boldsymbol A}_x(x|\lambda)\bigg)\,,
\ee
obeys 
\be\label{PBrel1}
\big\{{\boldsymbol M}(\lambda_1) \begin{array}{ccc} \\[-0.4cm] \otimes \\[-0.35cm] , \end{array}{\boldsymbol M}(\lambda_2)\,\big\}=
\big[{\boldsymbol M}(\lambda_1)\,{\otimes}\, {\boldsymbol M}(\lambda_2), {\boldsymbol r}(\lambda_1/\lambda_2)\,\big]\ ,
\ee
which can be thought of as the classical limit of  eq.\eqref{RTT1}
with ${\boldsymbol r}(\lambda)$  being the classical counterpart to the $R$-matrix.
The  Poisson algebra  \eqref{PBrel1}  is key
 in the Hamiltonian treatment 
of the integrable field theory as it immediately
implies the existence of a commuting family of conserved charges
generated by the trace of the monodromy  matrix.
\bigskip

To see how\,\eqref{PB0} leads to the classical Yang-Baxter
Poisson algebra \eqref{PBrel1},
one can  discretize  the path-ordered  integral  in \eqref{mod2}
 on a finite number of segments so that ${\boldsymbol M}(\lambda)$  is given by
an ordered product of elementary transport matrices ${\boldsymbol M}_n={\overset{\leftarrow}{\cal P}}
  \exp\big(\int_{x_n}^{x_{n+1}}{\rm d} x\,{\boldsymbol  {A}}\big)$. 
Since the r.h.s. of \eqref{PB0}  is proportional to the $\delta$-function, the Poisson brackets of  ${\boldsymbol M}_n$ 
corresponding to different
segments of the path vanish.  Then  the proof of eq.\,\eqref{PBrel1} becomes  practically equivalent to 
the ``lattice derivation'' of the quantum relation \eqref{RTT1} pictured in fig.\,\ref{fig1}.

\bigskip

For many interesting integrable systems, 
 the Poisson brackets
of the flat connection are ``non-ultralocal'': they
are modified from \eqref{PB0} by the presence of a term proportional to $\delta'(x-y)$. 
This results in ambiguities 
in the calculation of the Poisson brackets of the monodromy matrix
which come from contact terms  arising from the integration of the derivative of the $\delta$-function.
In the work  \cite{Maillet:1985ek}
a certain ``equal-point'' limiting prescription was put forward to handle such ambiguities which 
enables the introduction of a commuting family of conserved charges.
However the fundamental relations \eqref{PBrel1}
are modified in this approach
and it is unclear how to proceed with the quantization of the model
even at the formal algebraic level. 
The natural question arises of whether
it is possible to find a way of handling the contact terms such that \eqref{PBrel1} is unchanged. 
For the case of the Principal Chiral Field such a procedure was 
proposed in the work \cite{Duncan:1989vg}. In these notes, we will tackle this question differently 
by using an explicit realization of the quantum Yang-Baxter algebra \eqref{RTT1} and taking
its classical limit. We'll discuss the implications of our results for the  two parameter deformation
of the $SU(2)$ Principal Chiral Field introduced in \cite{Fateev:1996ea}.

\vskip 0.3in

\section{From quantum universal $R$-matrix to  $U(1)$ current algebra realization of Yang-Baxter Poisson structure}

\bigskip
The algebraic structure underlying eq.\,\eqref{RTT1} 
was clarified within the theory of quasi-triangular Hopf algebras by Drinfeld  \cite{Drinfeld1986}.
A basic example
is when the r$\hat{{\rm o}}$le of the Hopf algebra is played by $U_q(\widehat{\mathfrak{g}})$ 
 -- the
quantum deformation of the universal enveloping algebra of
the affine algebra \cite{Drinfeld1986,Jimbo:1985zk}. 
In this case a 
crucial element is  the universal $R$-matrix which lies in the
tensor product $U_q(\widehat{\mathfrak{g}})\otimes U_q(\widehat{\mathfrak{g}})$
and satisfies the relation  
\be\label{YB1}
{\cal R}^{12}\,{\cal R}^{13}\,{\cal R}^{23}\,=\,{\cal R}^{23}\,{\cal R}^{13}\,{\cal R}^{12}\ .
\ee
An important feature of ${\cal R}$ is that it is decomposed as
 ${\cal R}\in U_q(\widehat{{\mathfrak b}}_+)\otimes U_q(\widehat{{\mathfrak b}}_-)$ where 
$U_q(\widehat{{\mathfrak b}}_\pm)$  stand for the Borel subalgebras of $U_q(\widehat{\mathfrak{g}})$.
If we consider now the evaluation homomorphism of $U_q(\widehat{{\mathfrak g}})$ to the loop algebra
$U_q({\mathfrak{g}})[\lambda,\lambda^{-1}]$ and specify an $N$-dimensional matrix representation $\pi$ of
 $U_q(\mathfrak{g})$, then 
 \bea\label{aissia}
{\boldsymbol L}(\lambda) = \big(\pi(\lambda)\otimes 1\big)[{\cal R}]\,,
\eea
 is a 
$U_q(\widehat{{\mathfrak b}}_-)$-valued 
$N\times N$ matrix
whose entries depend on an auxiliary parameter $\lambda$. 
In its turn, the formal algebraic relation \eqref{YB1}  becomes the Yang-Baxter algebra \eqref{RTT1} 
with ${\boldsymbol M}$ substituted by ${\boldsymbol L}$ 
while
\bea
{\boldsymbol R}(\lambda_2/\lambda_1)=\big(\pi(\lambda_1)\otimes 
\pi(\lambda_2)\big)[{\cal R}]\, .\nonumber
\eea

These notes will mostly focus on ${\mathfrak g}=\mathfrak{ sl}_2$. In this case,
the Borel subalgebra $U_q(\widehat{{\mathfrak b}}_+)$  is generated by four elements,
$\{y_0,y_1,h_0,h_1\}$ and its evaluation homomorphism is defined by
\be\label{evalhomo1}
y_0\mapsto \lambda\, q^{-\frac{{\tt h}}{2}} \,{\tt e}_+\ , \qquad y_1\mapsto \lambda \,q^{\frac{{\tt h}}{2}} \,
{\tt e}_-\ , \qquad h_0\mapsto {\tt h}\ , \qquad h_1\mapsto -{\tt h}\ ,
\ee
where $\{ {\tt h},{\tt e}_\pm\}$ are the generators of $U_q({\mathfrak{sl}}_2)$, 
subject to the commutation relations
\be\label{comm5}
[{\tt h},\,{\tt e}_\pm]=\pm\, 2\,{\tt e}_\pm \ , \qquad [{\tt e}_+,\,{\tt e}_-]=\frac{q^{{\tt h}}-q^{-{\tt h}}}{q-q^{-1}} \ .
\ee
Below, with some abuse of notation,
we will not distinguish between the formal generators of $U_q({\mathfrak{sl}}_2)$ and their
matrices in a finite
dimensional representation. 
Explicitly, using the formula for the universal $R$-matrix
given in \cite{Khoroshkin:1994um}, 
one can obtain ${\boldsymbol L}(\lambda)$ as a formal series expansion in 
powers of the spectral parameter $\lambda$,\footnote{
In fact, eq.\,\eqref{series1} follows from
an expression of the $R$-matrix which 
is equivalent to the one in \cite{Khoroshkin:1994um} (and used in \cite{Bazhanov:1998dq})
upon the substitution $q\mapsto q^{-1}$ (see eq.\,\eqref{qdef}).
This is to keep with the conventions of the recent work \cite{Bazhanov:2017nzh}.}
\bea\label{series1}
&&\hspace{-1.2cm}{\boldsymbol L}(\lambda)=\bigg[1+\lambda\,(q-q^{-1})\,
(x_0\,q^{\frac{{\tt h}}{2}}\,{\tt e}_++x_1\,q^{-\frac{{\tt h}}{2}}\,{\tt e}_-)
+\lambda^2\, \frac{(q-q^{-1})^2}{1+q^2}\,
\bigg(
\,x_0^2\ (q^{\frac{{\tt h}}{2}}\,{\tt e}_+)^2+
x_1^2\ (q^{-\frac{{\tt h}}{2}}\,{\tt e}_-)^2\nonumber\\[0.3cm]
&&\hspace{-0.5cm}
+\,\frac{q^2\,x_0x_1-x_1x_0}{1-q^{-2}}\ (q^{\frac{{\tt h}}{2}}\,{\tt e}_+)
(q^{-\frac{{\tt h}}{2}}\,{\tt e}_-)+
\frac{q^2\,x_1x_0-x_0x_1}{1-q^{-2}}\ (q^{-\frac{{\tt h}}{2}}\,{\tt e}_-)
(q^{\frac{{\tt h}}{2}}\,{\tt e}_+)\bigg)+\ldots\bigg]
q^{-\frac{1}{2}{\tt h}\, h_0}\,. 
\eea
The expression in the square brackets contains only the generators $x_0,x_1\in U_q(\widehat{{\mathfrak b}}_-)$ 
satisfying the Serre relations
\be\label{Serre1}
x_i^3x_j-[3]_q\,x_i^2x_jx_i+[3]_q\,x_ix_jx_i^2-x_jx_i^3=0 \qquad (i,j=0,1) \, ,
\ee
where $[n]_q\equiv(q^n-q^{-n})/(q-q^{-1})$. 
Note that the two remaining generators  $h_0,h_1$, which obey
\be\label{comm2}\arraycolsep=0.5cm
\begin{array}{lll}
  [h_0,x_0]=-[h_1,x_0]=-2x_0\, , & [h_0,x_1]=- [h_1,x_1]=2x_1\,, & [h_0,h_1]=0 \, ,
\end{array}
\ee
appear only in an overall factor multiplying the square bracket $[\,\ldots\,]$ in \eqref{series1}. In fact, since $h_0+h_1$
is a central element, for our purposes and
without loss of generality we have set it to be zero. 
\bigskip

Until this point
there was no need to specify a representation of  $U_q(\widehat{{\mathfrak b}}_-)$ -- 
the Yang-Baxter relation \eqref{RTT1} is satisfied identically 
provided \eqref{Serre1},\,\eqref{comm2} hold true. 
In ref.\cite{Bazhanov:1998dq}, a 
representation of $U_q(\widehat{{\mathfrak b}}_-)$ was considered 
in the (extended) Fock space of a single bosonic field. 
The Borel
generators $x_0,\,x_1$ were given by the integral expressions
\be\label{contourInt1}
x_0 = \frac{1}{q-q^{-1}}\, \int_0^R{\rm d}z\, V^+(z) \ ,\qquad x_1 = \frac{1}{q-q^{-1}}\,\int_0^R{\rm d}z\, V^-(z) \ .
\ee
Here the vertex operators
\be\label{vertex1}
V^\pm(z)=\re^{\mp 2\ri\beta\varphi}(z)\ \nonumber
\ee
are built from the bosonic field
\be\label{phidef1}
\varphi(z)=\varphi_0+\frac{2\pi z}{R}\,\hat{p}+\ri\,\sum_{n\ne 0} \frac{a_n}{n}\,\re^{-\frac{2\pi\ri n}{R}z}
\ee
whose Fourier coefficients satisfy the commutations relations of the Heisenberg algebra
\be\label{phidef2}
[a_n,a_m]=\,\tfrac{n}{2}\,\delta_{n+m,0}\ , \qquad [\varphi_0,\hat{p}]=\tfrac{\ri}{2}\ .
\ee
The remaining generator $h_0=-h_1$ coincides with the 
zero mode momentum $\hat{p}$ up to a simple factor:
\be\label{h0gen}
 h_0=\frac{2}{\beta}\,\hat{p} \ .
\ee
The parameter $\beta$ appearing in the above formulae is related to the deformation parameter $q$ as
\be\label{qdef}
q=\re^{-\ri\pi\beta^2} \ . 
\ee
Defining the Fock space ${\cal F}_p$ as 
the highest weight module of the Heisenberg algebra with highest weight vector
$| p\rangle$:
$
\hat{p}\,| p\rangle = p\,| p\rangle \nonumber
$,
it easy to see that the generators \eqref{contourInt1} act as
\be
x_0 :\ \  {\cal F}_p\mapsto {\cal F}_{p-\beta}\ , \ \ \ \ \ \ \ \ x_1 :\ \  {\cal F}_p\mapsto {\cal F}_{p+\beta}  \nonumber
\ee
and hence that the matrix elements of ${\boldsymbol L}(\lambda)$ \eqref{series1} are operators in the extended Fock space
$\oplus_{n=-\infty}^\infty {\cal F}_{p+ n\beta}$. 
\bigskip

It was observed in \cite{Bazhanov:1998dq} that
using the commutation relations, 
\be\label{Vcomm1}
V^{\sigma_1}(z_1)\,V^{\sigma_2}(z_2)=q^{2\sigma_1\sigma_2}\,
V^{\sigma_2}(z_2)\,V^{\sigma_1}(z_1) \, , \ \ \ \ \ z_2>z_1 \ \ \ \ \ (\sigma_{1,2}=\pm)
\ee
the monomials built from the generators $x_0,\,x_1$ can be expressed in terms of the ordered integrals
\be\label{Iordered}
J(\sigma_1,\ldots,\sigma_m) = \!\!\!\!\!\!\!\!\int\limits_{R> z_1> z_2>\ldots> z_m> 0}
\hspace{-1cm}{\rm d}z_1\ldots{\rm d}z_m\, V^{\sigma_1}(z_1)\ldots V^{\sigma_m}(z_m)\ ,
\ee
which yields the following expression for ${\boldsymbol L}(\lambda)$
\be\label{Mseries}
{\boldsymbol L}(\lambda)=\,
\sum\limits_{m=0}^\infty\ \lambda^m\!\!\!\sum\limits_{\sigma_1\ldots\sigma_m=\pm}
\big(q^{\frac{{\tt h}}{2}\sigma_1}{\tt e}_{\,\sigma_1}\big)\ldots 
\big(q^{\frac{{\tt h}}{2}\sigma_m}{\tt e}_{\,\sigma_m}\big)\ J(\sigma_1,\ldots,\sigma_m)\ \re^{\ri\pi\beta\, \hat{p}\,{\tt h}}\ .
\ee
The latter is recognized as the path ordered exponent
\be\label{Mordered1}
{\boldsymbol L}(\lambda)=\ \overset{\leftarrow}{{\cal P}}\exp\bigg(\lambda\!\int_0^R{\rm d}z\ \Big(V^+\,q^{\frac{{\tt h}}{2}}\,{\tt e}_++V^-\,
q^{-\frac{{\tt h}}{2}}\,{\tt e}_-\Big)\bigg)\,\re^{\ri\pi\beta\, \hat{p}\,{\tt h}}\ .
\ee
It should be emphasized that since
the OPE of the vertex operators is singular,
\be
V^\pm(z_2)\,V^\mp(z_1)\big|_{z_2\to z_1+0}\sim(z_2-z_1)^{-2\beta^2}\ , \nonumber
\ee
the ordered integrals
are well defined only for $
0<\beta^2<\tfrac{1}{2} $.
However, 
each term in the formal series expansion \eqref{series1},  
being expressed in terms of the basic contour integrals
$x_0,\,x_1$, is well defined for all values of $\beta$ except 
the cases when $\beta^2=1-\frac{1}{2n}$ with
$
n=1,2,3,\ldots\ . 
$
In fact, the series expansion \eqref{series1} can be thought of as an analytic regularization
of the divergent path-ordered exponent \eqref{Mordered1}
within the domain $\frac{1}{2}<\beta^2<1$.
\bigskip

Let's consider the classical limit
where $\beta\to 0$ so that the deformation parameter $q$ tends to one.
The commutation relations \eqref{comm5} 
turn into
\be\label{comm4}
[{\tt h},{\tt e}_\pm]= \pm 2{\tt e}_\pm \ , \qquad [{\tt e}_+,{\tt e}_-]={\tt h}\ ,
\ee
while 
$
\phi\equiv \beta\,\varphi\ 
$
becomes a classical quasiperiodic field,
\be\label{classfield1}
\phi(R)-\phi(0)=2\pi P\,,
\ee
satisfying the Poisson bracket relations
\be\label{classfield3}
\big\{\phi(z_1),\phi(z_2)\big\}=-\tfrac{1}{4}\,\epsilon(z_1-z_2)
\ee
with $\epsilon(z)=2m+1$ for $mR<z<(m+1) R\ \ (m\in \mathbb{Z})$.
Since for small $\beta$ there is no convergence issue
the $\beta\to 0$ limit of \eqref{Mordered1} is straightforward, yielding
 the classical path-ordered exponent of the form
\be\label{Climit1}
{\boldsymbol L}_{\rm cl}(\lambda)=
\overset{\leftarrow}{{\cal P}}\exp\bigg(\lambda\int_0^R{\rm d}z\,
\big(\re^{-2\ri\phi}\ {\tt e}_++\re^{2\ri\phi}\ {\tt e}_-\big)\bigg)\,\re^{\ri\pi P\, {\tt h}}\ . \nonumber
\ee
Here, abusing notation for the sake of readability,
we denote the classical counterparts to the quantum operators 
by the same symbols, in particular,
${\tt e}_\pm$ now fulfill relations \eqref{comm4}
 and $\phi$ is the classical field 
satisfying \eqref{classfield1},\,\eqref{classfield3}.

\bigskip
The matrix ${\boldsymbol L}_{\rm cl}(\lambda)$ 
essentially coincides with the monodromy matrix for the linear differential equation
\be\label{LDop1}
\big(\partial_z-{\boldsymbol A}\big){\boldsymbol\Psi}(z)=0\ ,
\ee
where
\be\label{mKdV1}
{\boldsymbol A}(z|\lambda)=j(z)\, {\tt h} + \lambda\,({\tt e}_++{\tt e}_-)\,, \qquad j(z)=\ri\,\partial\phi(z)\,.
\ee
Indeed, a simple calculation leads to
\be\label{Mres1}
 {\boldsymbol L}_{\rm cl}(\lambda)\,\re^{\ri\pi P\, {\tt h}}=
{\boldsymbol \Omega}^{-1}\ \bigg[\
 \overset{\leftarrow}{{\cal P}}\exp\bigg(\, {\int_0^R}{\rm d}z\,{\boldsymbol A}(z|\lambda)\bigg)\bigg]\
{\boldsymbol \Omega}
\ee
with  ${\boldsymbol \Omega}=\re^{\ri\phi(R){\tt h}}$. We now observe that
the ordinary differential equation \eqref{LDop1} is the  auxiliary linear problem for the classically integrable mKdV hierarchy, while
\be\label{mKdV2}
\big\{j(z_1),j(z_2)\big\}=-\tfrac{1}{2}\,\delta'(z_1-z_2) \, 
\ee
(which follows from \eqref{classfield3}) is its first Hamiltonian structure.
The above formula implies that the Poisson brackets of ${\boldsymbol A}$ do not have the
ultralocal form \eqref{PB0} and, as it was mentioned earlier, the computation of
the Poisson brackets for the path-ordered exponent 
$\overleftarrow{{\cal P}}\exp\big(\int_0^R{\rm d}z\,{\boldsymbol A}\,\big)$
is inevitably met with ambiguities in treating the contact terms. Nonetheless,
 the classical limit of the Yang-Baxter algebra \eqref{RTT1} unambiguously
yields that 
 \eqref{PBrel1} is satisfied with ${\boldsymbol M}(\lambda)$ substituted by  ${\boldsymbol L}_{\rm cl}(\lambda)$
from \eqref{Mres1}
while ${\boldsymbol r}(\lambda)={\boldsymbol r}_-(\lambda)$,
where
\be\label{CRmatrix111}
{\boldsymbol r}_-(\lambda) = -\frac{1}{\lambda-\lambda^{-1}}\,\Big(\,{\tt e}_+\otimes {\tt e}_-+{\tt e}_-\otimes {\tt e}_++
                                                 \tfrac{1}{4}(\lambda+\lambda^{-1})\,{\tt h}\otimes {\tt h}\,\Big)\ .
\ee
Thus we see that 
starting from an explicit realization of the quantum algebra \eqref{RTT1}
and taking the classical limit
is a clear-cut way of obtaining
the monodromy matrix satisfying the classical Yang-Baxter Poisson algebra
 for a 
non-ultralocal flat connection.

\vskip 0.3in

\section{From quantum universal $R$-matrix to  $SU(2)$ current 
algebra realization of Yang-Baxter Poisson structure}\label{sec3}

\bigskip
It is known \cite{Feigin:2001yq,Bazhanov:2013cua}  that the Borel subalgebra 
$           U_q(\widehat{{\mathfrak b}}_-)
\subset U_q(\widehat{\mathfrak{sl}}_2)$
admits a realization 
with $x_0$ and $x_1$ given by
  \eqref{contourInt1},
where the vertices $V^{\pm}$ are built from three bosonic fields $\varphi_1$, $\varphi_2$, $\varphi_3$:
\be\label{vertex2}
V^{\pm}=\frac{1}{2b^2}\,\big( \ri b\,\partial\varphi_3+ \alpha_2\,\partial\varphi_2\pm\alpha_1
\,\partial\varphi_1\,\big)\re^{\pm\frac{\varphi_3}{b}}\ .
\ee
The expansion coefficients of $\varphi_i$, defined by the formula similar to \eqref{phidef1},
generate three independent copies of the Heisenberg algebra \eqref{phidef2}. 
The relation \eqref{h0gen} is replaced now by
\bea\label{aisi}
h_0=-h_1=-4\ri b\,\hat{p}_3\ ,
\eea
where $\hat{p}_3$ is the zero mode momentum of the field $\varphi_3$.
It should be highlighted that the parameters $\alpha_1$, $\alpha_2$, $b$ appearing in 
eq.\,\eqref{vertex2} are subject to the constraint
\be\label{constraint1}
\alpha_1^2+\alpha_2^2-b^2=\tfrac{1}{2}
\ee
and $b$ is related to the deformation parameter $q$ as
\be\label{oioqiwqw}
q=\re^{\frac{\ri\hbar}{2}}\  \ \ \ \ \ {\rm with} \ \ \ \ \ \hbar =\frac{\pi}{2b^2}\, .
\ee

\bigskip

The set of  generators  $\{x_0, x_1, h_0, h_1\}$  defined by \eqref{contourInt1},\,\eqref{vertex2},\,\eqref{aisi} 
fulfill the Serre 
and
 commutation relations \eqref{Serre1},\,\eqref{comm2}.
In consequence, 
${\boldsymbol L}(\lambda)$\ \eqref{aissia}
derived from the universal $R$-matrix by taking this realization of $U_q(\widehat{{\mathfrak b}}_-)$
satisfies the Yang-Baxter algebra \eqref{RTT1}.
The formal power series expansion in $\lambda$ \eqref{series1} is still applicable however eq.\,\eqref{Mseries}, which expresses
${\boldsymbol L}(\lambda)$ in terms of the ordered integrals,   
turns out to be problematic because of an issue with convergence.
Indeed, the OPE
\be
V^{\sigma_2}(z_2)\,V^{\sigma_1}(z_1)\sim(z_2-z_1)^{- 2-\sigma_1\sigma_2/(2b^2)} \ \ \ \ \ \ \ \ \ \ \  (\sigma_{1,2}=\pm) \nonumber
\ee
 is more singular now and the ordered integrals \eqref{Iordered} in general diverge.
Thus 
the path ordered exponent expression for ${\boldsymbol L}(\lambda)$ \eqref{Mordered1} 
that was obtained 
from recasting the contour integrals into the ordered integrals using the commutation relations \eqref{Vcomm1} (which are still valid)
is ill defined.
When taking the classical limit $b\to \infty$ it is essential to keep this in mind.

\bigskip
To study the classical limit, it is convenient to work with $\phi_i\equiv\varphi_i/(2b)$ which become classical
quasi-periodic fields 
\be\label{classfield2}
\phi_i(R)-\phi_i(0)=2\pi P_i\ \ \ \ \ \ (i=1,2,3)
\ee
satisfying equations similar
 to \eqref{classfield3}. 
As it follows from \eqref{contourInt1},\,\eqref{vertex2},\,\eqref{constraint1}
the classical counterparts of $x_0$ and $x_1$
are given by 
\be\label{chiclass}
\chi_0=\lim\limits_{b\to \infty}\,(q-q^{-1})\,x_0=\int_0^R{\rm d}z\,V^+_{{\rm cl}}(z) \ , \ \ \ \ \ \
 \chi_1=\lim\limits_{b\to \infty}\,(q-q^{-1})\,x_1=\int_0^R{\rm d}z\,V^-_{{\rm cl}}(z)\ ,
\ee
where
\be\label{cvertex1}
V^\pm_{{\rm cl}}=\big(\ri\,\partial\phi_3
+\tfrac{1}{\sqrt{1+\nu^2}}\ \partial\phi_2
\pm\tfrac{\nu}{\sqrt{1+\nu^2}}\ \partial\phi_1
\big)\,\re^{\pm 2\phi_3}
\ee
and
 \be
 \nu\equiv\lim\limits_{b\to\infty}\alpha_1/\alpha_2\,. \nonumber
 \ee 
Since the expression \eqref{series1} for ${\boldsymbol L}(\lambda)$ does not have problems
 with convergence, we will use it for taking the classical limit.
 Each term in the series  \eqref{series1} is a polynomial w.r.t. the non-commutative variables
 $x_0$ and $x_1$ with coefficients depending on the deformation parameter $q$.
To take the $\hbar\to 0$ limit
one should expand $q$ \eqref{oioqiwqw}
for small $\hbar$, express the result in terms of commutators and then replace
the commutators with Poisson brackets using 
 the correspondence principle $ [\,.\,,\,.\,]\mapsto \ri\hbar\ \{\,.\,,\,.\,\}$.
It is easy to see that with this procedure the first few terms shown in \eqref{series1} become
 \bea\label{Cseries1}
&&\hspace{-1cm}\lim\limits_{\hbar\to 0}{\boldsymbol L}(\lambda)=\bigg[1+
\lambda\,
(\chi_0\,{\tt e}_++\chi_1\,{\tt e}_-\,)\,+\\[0.3cm]
&&\hspace{-1cm}\,\hspace{0.0cm}\tfrac{1}{2}\,\lambda^2\,\Big(
\chi_0^2\ {\tt e}_+^2+\chi_1^2\  {\tt e}_-^2 
+\big(\chi_0\chi_1+\{\chi_0,\chi_1\}\big)\ {\tt e}_+{\tt e}_-+
\big(\chi_0\chi_1+\{\chi_1,\chi_0\}\big)\ {\tt e}_-{\tt e}_+\Big)+\ldots\ \bigg]
\re^{-\pi P_3\, {\tt h}}\nonumber
\eea
where ${\tt h},{\tt e}_\pm$ satisfy
 the commutation relations of the $\mathfrak{sl}_2$ algebra \eqref{comm4}.
\bigskip

 The calculation for higher order coefficients quickly becomes cumbersome. 
For example, the formal expansion of 
${\cal R}\,q^{\frac{h_0\otimes h_0}{2}}\in U_q(\widehat{{\mathfrak b}}_+)\otimes U_q(\widehat{{\mathfrak b}}_-)$
contains the term $y_1y_0^2y_1\otimes P_4^{(1001)}(x_0,x_1)$ with
\be
P_4^{(1001)}(x_0,x_1)=\frac{q^6(q-q^{-1})^2}{\,[4]_q\,[2]_q}\ 
\Big(x_0^2x_1^2-[3]_q\,x_0x_1x_0x_1+x_0x_1^2x_0+[3]_q\,x_1x_0^2x_1-[3]_q\,x_1x_0x_1x_0+x_1^2x_0^2\,\Big) \nonumber
\ee
which makes a fourth order contribution to the series   \eqref{series1} once the evaluation homomorphism \eqref{evalhomo1} 
of $y_0$, $y_1$ is taken.
Expanding $q$ for small $\hbar$ in $P_4^{(1001)}(x_0,x_1)$ yields
\bea
&&\hspace{-0.5cm}P_4^{(1001)}(x_0,x_1)=-\tfrac{1}{8}\,\hbar^2\,\Big(1+
O(\hbar)\Big)\times\nonumber \\[0.3cm]
&&\hspace{-0.5cm}\Big([x_0,[x_0,x_1]]x_1+x_1[x_0,[x_0,x_1]]-[x_0,x_1]^2+\hbar^2
\big(x_0x_1x_0x_1+x_1x_0x_1x_0-x_1x_0^2x_1\big)\,+O(\hbar^4)\,\nonumber
\Big) \,.
\eea
Now, replacing $x_0$, $x_1$ by their classical counterparts \eqref{chiclass}, using 
the correspondence principle and taking the limit $\hbar\to 0$ gives
\be
\lim\limits_{\hbar\to 0} P_4^{(1001)}(x_0,x_1) = 
\tfrac{1}{8}\,\big(2\chi_1\{\chi_0,\{\chi_0,\chi_1\}\}-\{\chi_0,\chi_1\}^2+\chi_0^2\chi_1^2\,\big)\, . \nonumber
\ee
For the full contribution to the fourth order of \eqref{Cseries1}
one should take into account
all sixteen polynomials $P_4^{(i_1i_2i_3i_4)}(x_0,x_1)$ with $i_1,i_2,i_3,i_4=0,1$ corresponding
to the terms $y_{i_1}y_{i_2}y_{i_3}y_{i_4}\otimes P_4^{(i_1i_2i_3i_4)}(x_0,x_1)$ in the expansion of the 
universal $R$-matrix.
\bigskip

Our calculations to fifth order in $\lambda$ support the existence of the limit
\be\label{Mlim}
\lim\limits_{\hbar\to 0}{\boldsymbol L}={\boldsymbol L}_{\rm cl}\ .
\ee
By construction, ${\boldsymbol L}_{\rm cl}$ is a formal series expansion in $\lambda$ whose 
coefficients are built from $\chi_0$, $\chi_1$ and their Poisson brackets.\footnote{%
Note  that the elements $\chi_0$ and $\chi_1$ satisfy the classical analogs of the Serre relations \eqref{Serre1},
$$
\{\chi_i,\,\{\chi_i,\,\{\chi_i,\,\chi_j\}\}\}=\chi_i^2\,\{\chi_i,\,\chi_j\}\, 
\qquad (i,j=0,1)\,.
$$
}
To proceed further, the latter need to be computed explicitly. 
This can be carried out along the following lines.
Starting from the relations
\be\label{PBfields1}
\big\{\phi_i(z_1),\,\phi_j(z_2)\big\}=-\tfrac{1}{4}\,\delta_{ij}\,\epsilon(z_1-z_2)\ 
\ee
it is easy to show that $V^\pm_{\rm cl}$ \eqref{cvertex1} and 
\be\label{cvertex0}
V^0_{{\rm cl}}\,=\,-2\,\big(\tfrac{1}{\sqrt{1+\nu^2}} \ \partial\phi_3-\ri\, \partial\phi_2\big)\, 
\ee
form a closed Poisson algebra
\begin{eqnarray}\label{PBalgebra1}
&&\hspace{-0.6cm}\{V^0_{{\rm cl}}(z_1),\,V^0_{{\rm cl}}(z_2)\}=-\frac{2\nu^2}{1+\nu^2}\ \delta'(z_1-z_2) \nonumber \\[0.2cm]
&&\hspace{-0.6cm}\{V^0_{{\rm cl}}(z_1),\,V^{\pm}_{{\rm cl}}(z_2)\}=\pm \,\frac{2}{\sqrt{1+\nu^2}}\ V^\pm_{{\rm cl}}(z_1)\,\delta(z_1-z_2) \\[0.2cm]
&&\hspace{-0.6cm}\{V^+_{{\rm cl}}(z_1),\,V^-_{{\rm cl}}(z_2)\}=-\frac{\nu^2}{1+\nu^2}\,\delta'(z_1-z_2)+
\frac{V^0_{{\rm cl}}(z_1)}{\sqrt{1+\nu^2}}\ \,\delta(z_1-z_2)+
V^+_{{\rm cl}}(z_1)\,V^-_{{\rm cl}}(z_2)\,\epsilon(z_1-z_2) \nonumber \\[0.2cm]
&&\hspace{-0.6cm}\{V^\pm_{{\rm cl}}(z_1),\,V^\pm_{{\rm cl}}(z_2)\}=-V^\pm_{{\rm cl}}(z_1)\,V^\pm_{{\rm cl}}(z_2)\,\epsilon(z_1-z_2) \ . \nonumber
\end{eqnarray}
Recall that $\chi_0$ and $\chi_1$ are given by integrals over the classical vertices \eqref{chiclass} so that these relations 
are sufficient for the explicit  calculation of any of the Poisson brackets occurring in the r.h.s of \eqref{Cseries1}. 
However, due to the presence of the derivative of the $\delta$-function in \eqref{PBalgebra1}, ambiguous
integrals occur in the computations.
For instance: $\{\chi_0,\chi_1\}=c_1\,\nu^2/(1+\nu^2)+\ldots\ $
with
\be\label{c1sing}
c_1=-\int_0^{R}{\rm d}z_1\,{\rm d}z_2\,\delta'(z_1-z_2)=\int_0^R{\rm d}z\,\big(\delta(z-R)-\delta(z)\big)\, .
\ee
In general, one is faced with many other sorts of integrals involving $\delta'(z_1-z_2)$.
However, they are not all independent and their number can be reduced 
if, before performing explicit calculations,
one uses the Jacobi identity and skew-symmetry 
 to bring the
Poisson brackets to the form
\be\label{basicPB}
\{\chi_{\sigma_1},\{\chi_{\sigma_2},\{\chi_{\sigma_3},\{\ldots,\{\chi_{\sigma_{m-1}},\chi_{\sigma_m}\}\ldots\}\  \ \  \ \ \ \ \ 
(\sigma_1,\ldots,\sigma_m=0,1)
\ee 
(e.g.,
$\{\{\chi_0,\chi_1\},\{\chi_0,\chi_1\}\}=
\{\chi_0,\{\chi_1,\{\chi_1,\chi_0\}\}\}+\{\chi_1,\{\chi_0,\{\chi_0,\chi_1\}\}\}
$).
This way, in our fifth order computations
we were met with only two more types of ambiguous integrals. The first is of the form
\begin{eqnarray}
I_1&=&\int_0^{R}{\rm d}z_1\ldots{\rm d}z_4\, \delta'(z_1-z_3)\,\epsilon(z_2-z_3)\,\epsilon(z_3-z_4)\ F(z_2)\,G(z_4)\ ,
\nonumber
\end{eqnarray}
where $F$ and $G$ are some functions.
Formal integration by parts w.r.t. $z_3$ yields 
\be\label{Aint1}
I_1=c_1\int_0^R{\rm d}z_1\,{\rm d}z_2\,F(z_1)\,G(z_2) \nonumber
\ee
with $c_1$ as in \eqref{c1sing}.
The other ambiguous integral is
\begin{eqnarray}
I_2&=&\int_0^{R}{\rm d}z_1{\rm d}z_2{\rm d}z_3\,F(z_2)\,\epsilon(z_2-z_3)\,\delta'(z_1-z_3)\, .
\nonumber
\end{eqnarray}
In this case, integration by parts leads to
\be\label{Aint2}
I_2=2\,(c_2-1)\int_0^{R}{\rm d}z\,F(z)\ \ \ \ \  {\rm with} \ \ \ \ \ \ c_2=\frac{1}{2}\,\int_0^{R}{\rm d}z\,
\big(\delta(z-R)+\delta(z)\big)\ .
\ee

We explicitly computed the expansion of 
${\boldsymbol L}_{\rm cl}$ to fifth order and found that all the ambiguities 
are absorbed in the two constants $c_1$ and $c_2$ \eqref{c1sing},\,\eqref{Aint2}.
Furthermore, if $c_1=0$ and $c_2$ is arbitrary, the series can be collected into a path-ordered exponent
\be\label{Lop3}
{\boldsymbol L}_{\rm cl}=
\overset{\leftarrow}{{\cal P}}\exp\bigg(\int_0^R{\rm d}z\, {\boldsymbol B}\bigg) \re^{-\pi P_3\, {\tt h}}
\ee
with
\begin{eqnarray}\label{eqA1}
{\boldsymbol B}\,&=&\,f\,\big({ V}^+_{\rm cl}(z)\ {\tt e}_+
+{ V}^-_{\rm cl}(z)\ {\tt e}_-\big)\,+\,
\tfrac{1}{2}\,g\, { V}^0_{\rm cl}(z)\, {\tt h}
\end{eqnarray}
and
\begin{eqnarray}\label{fgeq1}
f&=&\lambda\,\sqrt{1+\nu^2}\,\Big(1+(1+\nu^2\,(c_2-1))\,\lambda^2+(1+4\nu^2(c_2-1)+2\nu^4(c_2-1)^2)\,\lambda^4+O(\lambda^6)\,\Big) \nonumber \\[0.4cm]
g&=&\lambda^2\,\sqrt{1+\nu^2}\,\Big(1+(2\nu^2\,(c_2-1)+1)\,\lambda^2+O(\lambda^4)\Big)\ . \nonumber
\end{eqnarray}
That $c_1$ \eqref{c1sing} vanishes  seems to be a natural requirement as,
in the problem at hand, the $\delta$-function should be understood as the formal
series $\frac{1}{R}\sum_{m=-\infty}^\infty \re^{\frac{2\pi\ri m}{R} z}$ and hence $\delta(z-R)=\delta(z)$.
Note that for the periodic $\delta$-function the constant $c_2$ in \eqref{Aint2} becomes
\be\label{c2sing}
c_2=\int_0^R{\rm d}z\,\delta(z)\, .
\ee

Unfortunately there is no proof that the limit \eqref{Mlim} exists
and can be represented by eq.\,\eqref{Lop3} and \eqref{eqA1} with some functions $f$ and $g$ -- this 
has  been checked perturbatively to fifth order only. 
However, if this is accepted as a conjecture then $f$ and $g$ should have the form
\be\label{fgeq2}
f=\frac{\rho\,\sqrt{1+\nu^2}}{1-\rho^2}\ , \qquad \qquad g=\frac{\rho^2\,\sqrt{1+\nu^2}}{1-\rho^2}\ ,
\ee
where $\rho=\rho(\lambda)$ solves the equation
\begin{equation}\label{func1}
\lambda=\frac{\rho\,(1-\rho^2)}{1-(1+(1-c_2)\,\nu^2)\,\rho^2}\ .
\end{equation}
This follows from an analysis of the simplest matrix element of ${\boldsymbol L}_{\rm cl}$ for which
 the series \eqref{Cseries1} can be obtained to all orders in $\lambda$.

To summarize, we expect that the limit \eqref{Mlim} exists and results in \eqref{Lop3},
where ${\boldsymbol B}$ is  given by
\be\label{Atilde1}
{\boldsymbol B}(z|\rho)=\frac{\sqrt{1+\nu^2}}{1-\rho^2}\  \Big( \rho\,\big(V^+_{\rm cl}(z)\ {\tt e}_++V^-_{\rm cl}(z)\ {\tt e}_-\big)
+\,
\tfrac{1}{2}\,\rho^2\,V^0_{\rm cl}(z)\, {\tt h}\Big)\,
\ee
and with $\rho=\rho(\lambda)$ defined through the relation \eqref{func1}.
By construction ${\boldsymbol L}_{\rm cl}$  must satisfy the classical  Yang-Baxter Poisson algebra,
\be\label{PBrel11111}
\big\{{\boldsymbol L}_{\rm cl}(\rho_1) \begin{array}{ccc} \\[-0.4cm] \otimes \\[-0.35cm] , \end{array}
{\boldsymbol L}_{\rm cl}(\rho_2)\,\big\}=
\big[{\boldsymbol L}_{\rm cl}(\rho_1)\,{\otimes}\, {\boldsymbol L}_{\rm cl}(\rho_2), {\boldsymbol r}(\lambda_1/\lambda_2)\,\big]
\ee
with $\rho_{1,2}=\rho(\lambda_{1,2})$  and\footnote{
Note that here the classical $r$-matrix differs from the one in \eqref{CRmatrix111} by an overall sign.
}
\be\label{CRmatrix}
 {\boldsymbol r}(\lambda)=+\frac{1}{\lambda-\lambda^{-1}}\,\Big(\,{\tt e}_+\otimes {\tt e}_-+{\tt e}_-\otimes {\tt e}_++
                                                 \tfrac{1}{4}(\lambda+\lambda^{-1})\,{\tt h}\otimes {\tt h}\,\Big)\, .
\ee

\bigskip

Eq.\,\eqref{PBalgebra1} implies that the Poisson brackets of  ${\boldsymbol B}$ \eqref{eqA1}
are not local in the sense that apart from the $\delta$-function and its derivative
they contain terms with the $\epsilon$-function.
 Nevertheless, a simple calculation shows that the Lie algebra valued $1$-form 
${\boldsymbol B}(z|\rho)$
is gauge equivalent to
\be\label{Anew}
{\boldsymbol A}(z|\rho)= \frac{\rho\, \sqrt{1+\nu^2}}{1-\rho^2}\,\big(j^+(z)\ {\tt e}_++j^-(z)\ {\tt e}_-\big)+\,
\frac{1}{2}\,\bigg(\frac{\rho^2\, \sqrt{1+\nu^2}}{1-\rho^2}+\xi\bigg)\,j^0(z)\,{\tt h}
\ee
and the fields
\bea
j^\pm&=&\,\big(\ri\,\partial\phi_3
+\tfrac{1}{\sqrt{1+\nu^2}}\ \partial\phi_2\pm\,\tfrac{\nu}{\sqrt{1+\nu^2}}\ \partial\phi_1
\big)\,\re^{\pm 2\xi(\phi_3+\ri\phi_2)}\ \nonumber  \\[0.2cm]
j^0&=&-2\,\big(\tfrac{1}{\sqrt{1+\nu^2}}\,\partial\phi_3-\ri\,\partial\phi_2\big)\,\nonumber
\eea
satisfy  the classical current algebra
\bea\label{PBcurrents1}
\big\{j^+(z_1),j^-(z_2)\big\}&=&-\frac{\nu^2}{1+\nu^2}\ \delta'(z_1-z_2)+j^0(z_1)\,\delta(z_1-z_2) \nonumber \\[0.2cm]
\big\{j^0(z_1),j^\pm(z_2)\big\}&=&\pm2\,j^\pm(z_1)\ \delta(z_1-z_2)  \\[0.2cm]
\big\{j^0(z_1),j^0(z_2)\big\}&=&-\frac{2\nu^2}{1+\nu^2}\ \delta'(z_1-z_2) \nonumber \\[0.2cm]
\big\{j^\pm(z_1),j^\pm(z_2)\big\}&=& 0\ .\nonumber
\eea
The constant $\xi$ in the above formulae is given by
\be
\xi=\frac{\sqrt{1+\nu^2}}{1+\sqrt{1+\nu^2}}\ . \nonumber
\ee
It follows from eq.\,\eqref{PBcurrents1} that the  $\epsilon$-function is not present in the 
Poisson brackets of ${{\boldsymbol A}}$ \eqref{Anew} so they are local, although not ultralocal.
In terms of the $1$-form ${{\boldsymbol A}}$,
eq.\eqref{Lop3} 
can be re-written as
\be\label{conn3}
 {\boldsymbol L}_{\rm cl}(\rho)\,\re^{((2\xi-1)P_3+2\ri\xi P_2)\pi{\tt h}}=
 {\boldsymbol \Omega}^{-1}\ \bigg[
 \overset{\leftarrow}{{\cal P}}\,\exp\bigg(\, {\int_0^R}{\rm d}z\,{\boldsymbol A\big(z|\rho\big)}\bigg)\bigg]\
{\boldsymbol \Omega}\ ,
\ee
where $\,{\boldsymbol \Omega}=\exp\big((\xi-1)\,\phi_3(R)\,{\tt h}+\ri\,\xi\,\phi_2(R)\,{\tt h}\big)$
and $P_i$  are  defined by eq.\,\eqref{classfield2}. The r.h.s. of \eqref{conn3} is
 the monodromy matrix
for the linear problem \eqref{LDop1} with ${\boldsymbol A}$ given by \eqref{Anew} and $\rho$ 
playing the r$\hat{{\rm o}}$le of the auxiliary spectral parameter.

\bigskip

Despite that
the 
Poisson brackets of the $1$-form ${\boldsymbol A}$ are non-ultralocal for $\nu\ne 0$,
${\boldsymbol L}_{\rm cl}(\rho)$ in
\eqref{conn3} obeys the classical Yang-Baxter Poisson algebra \eqref{PBrel11111}.
The $\delta'$-terms
 introduce an ambiguity in
taking the classical limit
which is manifest in the arbitrary constant $c_2$ \eqref{c2sing}.
The effect of this is observed in the  finite renormalization of the 
spectral parameter $\lambda\mapsto \rho(\lambda)$ \eqref{func1}.
Notice that for the ultralocal case, i.e., $\nu=0$, the dependence on $c_2$
drops out and $\rho=\lambda$.

\section{Some facts about the Klim\v{c}\'{i}k  model\label{sec4}}
The Principal Chiral Field (PCF) is one of the keystone models of  integrable  field theory in 1+1 dimensions.
In the simplest setup, where the model is associated with
a simple Lie algebra ${\mathfrak g}$ equipped with the Killing form $\langle.\,,.\rangle$,
the Lagrangian is given by
\bea\label{oaspsaop0}
{\cal L}_{\rm PCF}=-\frac{4}{{\rm g}^2}\ \Big\langle\,{\boldsymbol U}^{-1}\partial_+ {\boldsymbol U},\,
{\boldsymbol U}^{-1}\partial_- {\boldsymbol U}\Big\rangle\, .
\eea
Here the field  ${\boldsymbol  U}(t,x)$ takes  values in  the Lie group ${\mathfrak G}$  corresponding to  the Lie algebra
so that ${\boldsymbol U}^{-1}\partial_\pm {\boldsymbol U}\in {\mathfrak g}$,
and the subscripts $\pm$ label the light-cone co-ordinates
\bea\label{apospsa}
x_\pm=t\pm x\ ,\ \ \ \ \ \ 
\partial_\pm={\textstyle\frac{1}{2}}\ (\partial_t\pm\partial_x)\ .
\eea

In Ref.\cite{Klimcik:2014bta},   Klim\v{c}\'{i}k introduced a two parameter deformation 
of the PCF. 
The construction uses the so-called Yang-Baxter operator --
a linear operator ${\hat {\boldsymbol R}}$  acting in
${\mathfrak g}$ which is defined through the 
root decomposition of the Lie algebra,
$
{\mathfrak g}={\mathfrak n}_+\oplus {\mathfrak h}\oplus {\mathfrak n}_-
$,
w.r.t. the Cartan subalgebra ${\mathfrak h}$.
Namely, for any element ${\tt  e}_\pm$ from the nilpotent subalgebras ${\mathfrak n}_\pm$:
${\hat {\boldsymbol R}}\big({\tt  e}_\pm\big)=\mp\ri\, {\tt  e}_\pm$,
while ${\hat {\boldsymbol R}}({\tt h})=0$ for $\forall \ {\tt  h}\in{\mathfrak h}$.
The Lagrangian of the Klim\v{c}\'{i}k  model with deformation parameters
$\varepsilon_1,\varepsilon_2$ is given by
\bea\label{oaspsaop}
{\cal L}_{\rm K}=-\frac{4}{{\rm g}^2}\ \Big\langle\,{\boldsymbol U}^{-1}\partial_+ {\boldsymbol U},\,
\big(\, {\hat  {\boldsymbol 1}}-\ri\varepsilon_1\,{\hat {\boldsymbol R}}_{\boldsymbol U}-
\ri\varepsilon_2\, {\hat {\boldsymbol R}}\,\big)^{-1} 
\big(
{\boldsymbol U}^{-1}\partial_- {\boldsymbol U}\big)\Big\rangle\, ,
\eea
where the action of
${\hat {\boldsymbol R}}_{\boldsymbol U}$  is defined as
\bea
{\hat {\boldsymbol R}}_{\boldsymbol U}({\tt a})={\boldsymbol U}^{-1}\, {\hat {\boldsymbol R}}
\big({\boldsymbol U}\,{\tt a}\,{\boldsymbol U}^{-1}\big)\, {\boldsymbol U}\ \ \ \ \ 
\ \ \ {\rm for}\ \ 
 \ \ \forall \,{\tt  a}\in{\mathfrak g}
\eea
(the symbol ${\boldsymbol U}\,(\ldots) \,{\boldsymbol U}^{-1}$  denotes the  adjoint action of the
group element ${\boldsymbol U}$ on  ${\mathfrak g}$). 
\bigskip

\subsection{Hamiltonian formulation\label{sec41}}
The Hamiltonian structure of the model can be described in terms of the currents
\bea\label{currentsK}
{\boldsymbol I}_\pm=
-2\ri\ \big({\hat {\boldsymbol 1}}\pm\ri \varepsilon_1\,{\hat {\boldsymbol R}}_{\boldsymbol U}
\pm\ri \varepsilon_2\,{\hat  {\boldsymbol R}}\big)^{-1}
\big({\boldsymbol U}^{-1}\partial_\pm  {\boldsymbol U}\big)\ .
\eea
A straightforward calculation yields that the Hamiltonian is given by 
\bea\label{hamK}
{ H}=\frac{1}{2{\rm g}^2}\int\rd x\, \big(\, 
\langle \, { { {\boldsymbol { I}}}}_+,\,{ { {\boldsymbol {  I}}}}_+\,\rangle+
\langle \, { { {\boldsymbol {  I}}}}_-,\,  { { {\boldsymbol {I}}}}_-\,\rangle\,\big)\,.
\eea
Starting from the Lagrangian \eqref{oaspsaop} one can show that the currents
$
{\boldsymbol I}_\pm=\sum_a I^a_\pm {\tt t}_a
$ \eqref{currentsK} 
obey the Poisson bracket relations
 \begin{equation}\label{ajsasysya}
{\rm g}^{-2}\,\big\{I^a_\sigma(x),\,I^b_{\sigma'}(y)\big\}\,=\,\sigma\,q^{ab}\delta_{\sigma\sigma'}\,\delta'(x-y)+
\sum\limits_{\sigma''}\,  F^{abc} (\sigma,\sigma'|\sigma'')\, q_{ cd}\,  I^d_{\sigma''}\,\delta(x-y)\,.
\end{equation}
Here the structure constants are given by
\bea\label{osasoasa}
2 F^{abc}(\pm\pm|\pm)&=&+(1+{\mathsf b})
 \ f^{abc}\pm  \ri { {\varepsilon}}_2\, \big({{\cal R}}^{c}{}_{d}\, f^{dba}+
 {{\cal R}^{b}}{}_{d}\, f^{dac}+{{\cal R}^{a}}{}_{d}\, f^{dcb}\, \big)\nonumber\\[0.2cm]
 2 F^{abc}(\pm \pm |\mp )&=&
- (1-{\mathsf b})\, f^{ab c}\pm   \ri { {\varepsilon}}_2
\, {{\cal R}^{c}}{}_{d}\,f^{dba} \\[0.2cm]
2F^{abc}(\pm \mp|\pm )&=&
+(1-{\mathsf b})\, f^{ab c}\mp   \ri { {\varepsilon}}_2
\, {{\cal R}^{b}}{}_{d}\, f^{dac}\nonumber\\[0.2cm]
2 F^{abc}(\mp \pm|\pm )&=&
+  (1-{\mathsf b})\, f^{ab c}\mp    \ri {{\varepsilon}}_2
\,{ {\cal R}^{a}}{}_d\, f^{dcb}\nonumber
\eea
with 
\bea\label{aoassi}
{\mathsf b}= \tfrac{1 }{2}\  (1+{ \varepsilon}_1^2-{ \varepsilon}_2^2)\ .\nonumber 
\eea
Also, ${{\cal R}}^{b}{}_{a}$ in the above formulae stands for the matrix elements of the Yang-Baxter operator
\bea
{\hat {\boldsymbol R}}({\tt t}_a)={\tt t}_b\, {{\cal R}^b}_a\ .\nonumber
\eea
\bigskip

In order to clarify the Poisson bracket relations \eqref{ajsasysya},
let us mention that 
 ${\boldsymbol I}_\pm$ are related by a linear transformation  to the currents
\bea\label{asisaisa}
{\boldsymbol J}_\pm(x)=\sum_a\,{ J}^a_\pm(x)\,{\tt t}_a \, , \ \ \ \ \ \ \ 
[{\tt t}_a, {\tt t}_b]=\ri\, {f_{ab}}^c\ {\tt t}_c\,,
\eea
which generate
two independent copies of the classical current algebra:
\begin{eqnarray}\label{jasssays}
 \big\{{J}^a_\sigma(x),\,{J}^b_{\sigma'}(y)\big\}\,=\,
\frac{1}{{\rm g}^2\varepsilon_1}\ \delta_{\sigma\sigma'}\, \sigma\,q^{ab}\,\delta'(x-y)+\delta_{\sigma\sigma'}\ 
{f^{abc}}\, q_{cd}\, 
{J}^d_\sigma(y)\,\delta(x-y)\,.
\end{eqnarray}
Here $\sigma,\sigma'=\pm$  and
\bea\label{iaosidosa}
q_{ab}=- \tfrac{1}{4}\, {f_{ac}}^d{f_{bd}}^c=\langle\, {\tt t}_a,{\tt t}_b\,\rangle\,.
\eea
For an explicit description of the  relation
  between ${\boldsymbol I}_\sigma$ and ${\boldsymbol J}_\sigma$ ($\sigma=\pm$),
it is convenient to use the root decomposition of the Lie algebra and represent  the currents in the form
\bea\label{jasysa}
{\boldsymbol I}_\sigma(x)={\boldsymbol I}^+_\sigma(x)+{\boldsymbol I}^0_\sigma(x)+{\boldsymbol I}^-_\sigma(x)\ :
\ \ \ \ {\boldsymbol I}^\pm_\sigma(x)\in {\mathfrak n}_\pm\ ,
\ \ \ \ \ {\boldsymbol I}^0_\sigma(x)\in {\mathfrak h}
\eea
and similarly for ${\boldsymbol J}_\pm$. Then it turns out that
\bea\label{aioasaso}
{\boldsymbol I}^A_{\sigma}=\frac{{\rm g}^2}{2}\sum_{\sigma'=\pm}
(1+\sigma\sigma'\,\varepsilon_1-A\,\sigma\varepsilon_2)\, {\boldsymbol J}^A_{\sigma'}\qquad\qquad {\rm with}
\qquad\qquad A=\pm,0\, .
\eea
Note that the Hamiltonian of the Klim\v{c}\'{i}k model \eqref{hamK} 
 is expressed in terms of the 
 currents ${\boldsymbol J}_\pm$ as
 \bea\label{jausuasa}
{ H}=\frac{{\rm g}^2}{4}\int\rd x\,\sum_{\sigma,\sigma'=\pm}\Big(\, A^\parallel_{\sigma\sigma'}\ \langle \,{\boldsymbol J}^0_\sigma,\, {\boldsymbol J}^0_{\sigma'}\,\rangle+
2 A^\perp_{\sigma\sigma'}\, \langle\, {\boldsymbol J}^+_\sigma,\, {\boldsymbol J}^-_{\sigma'}\,\rangle\, \Big)\ ,
\eea
where
 \bea\label{jassysa}
&&A^\parallel_{\pm\pm}=1+\varepsilon^2_1\ ,\ \ \ \ \ \ \ \ \ \ \ \ \ \ \ \ \,
A^\parallel_{\pm\mp}=1-\varepsilon^2_1\,,\nonumber
\\[0.2cm]
&&A^\perp_{\pm\pm}=1+\varepsilon_1^2-\varepsilon_2^2\, ,\ \ \ \ \ \ \ \ \ \ \ 
A^\perp_{\pm\mp}=(1+{\varepsilon}_1\mp 
{ \varepsilon}_2)(1-{\varepsilon}_1\pm {\varepsilon}_2)\ .
\eea
\bigskip

\subsection{Classical integrability}

A remarkable feature of the two parameter deformation of the PCF \eqref{oaspsaop} is that it preserves the 
integrability of the original model \cite{Klimcik:2014bta}. The flat connection appearing in the zero curvature representation
\be\label{ZCR1}
\big[\partial_+-{\boldsymbol A}_+,\,\partial_--{\boldsymbol A}_-\big]=0
\ee
is expressed in terms of the currents as
\bea\label{siqjkhdsjh2}
{\boldsymbol A}_\sigma=-\ \frac{\ri\,\varepsilon_2}{1-\rho_\sigma^2}\ \Big( 
  (\rho_\sigma)^{1-\sigma } \ {\boldsymbol I}^+_\sigma+ (\rho_\sigma)^{1 +\sigma }\ 
 {\boldsymbol I}^-_\sigma
 +{\textstyle \frac{1}{2}}\  \big(1+\rho_\sigma^2\big)\,{\boldsymbol I}^0_\sigma
\, \Big)\ \ \ \ \ \  \ \ (\sigma=\pm )\ ,
\eea
where the auxiliary parameters $\rho_\pm^2$ are subject to the single constraint\footnote{Eq.\,(20) 
from ref.\cite{Klimcik:2014bta} 
is equivalent to \eqref{siqjkhdsjh2} with ${\boldsymbol L}_\pm^{\alpha,\beta}(\zeta)={\boldsymbol A}_\pm$ provided the following
identifications are made $\alpha=\ri\,\varepsilon_1,\,  \beta=\ri\,\varepsilon_2$ and 
 the spectral parameter $\zeta=\frac{\rho_+^{2}+\rho_-^{-2}-2}{\rho_+^{2}-\rho_-^{-2}}$.} 
 \bea\label{hassyayt}
(\rho_+\rho_-)^2=
\frac{(1+\varepsilon_1-\varepsilon_2)(1-\varepsilon_1-\varepsilon_2)}{(1-\varepsilon_1+\varepsilon_2)
(1+\varepsilon_1+\varepsilon_2)}\ .
 \eea
  For our purposes, we will also use a slightly different gauge ${\boldsymbol A}^{(\omega)}_\pm $ which is defined as follows. 
 The equations of motion imply the conservation of the current ${\boldsymbol I}^0_\sigma$,\footnote{In 
the limit $\rho_+\to\infty$ and $\rho_-\to 0$ 
the connection 
\eqref{siqjkhdsjh2} becomes upper triangular, ${\boldsymbol A}_\sigma\in {\mathfrak n}_+\oplus {\mathfrak h}$, so that
eq.\,\eqref{conserve1} immediately follows from the zero curvature representation.}
\be\label{conserve1}
\partial_+\,{\boldsymbol I}^0_-+\partial_-{\boldsymbol I}^0_+=0 \, ,
\ee
which allows one to introduce the dual field ${\boldsymbol \omega}$
\be\label{woeiowei}
\partial_+{\boldsymbol \omega}=-\tfrac{1}{2}\,\varepsilon_2\,{\boldsymbol I}^0_+\,, \ \ \ \ \ \qquad
\partial_-{\boldsymbol \omega}=\tfrac{1}{2}\,\varepsilon_2\,
{\boldsymbol I}^0_-
\,,
\ee
taking  values in the Cartan subalgebra ${\mathfrak h}$. Then,
\be\label{oiqwoieoiw}
\partial_\pm-{\boldsymbol A}^{(\omega)}_\pm = 
\re^{+\ri{\boldsymbol \omega}}\,\big(\partial_\pm-{\boldsymbol A}_\pm\big)\, \re^{-\ri{\boldsymbol \omega}}\,  .
\ee

Apart from the local integrability condition -- the zero curvature representation --
proper global requirements should be imposed to ensure integrability of the model.
We consider the Klim\v{c}\'{i}k  model with the space co-ordinate restricted to the segment $x\in[\,0,R\,]$.
Since the Lagrangian \eqref{oaspsaop}
 is invariant under 
the transformation 
${\boldsymbol U}\mapsto {\boldsymbol H}_1{\boldsymbol U}{\boldsymbol H}_2$
where ${\boldsymbol H}_1,{\boldsymbol H}_2$ are elements of the Cartan subgroup
 ${\mathfrak H}\subset{\mathfrak G}$, a natural choice for the boundary conditions is
\be\label{asklas}
{\boldsymbol U}(t,x+R)={\boldsymbol H}_1{\boldsymbol U}(t,x){\boldsymbol H}_2\,,\ \ \ \ \ \ \ 
{\boldsymbol H}_1,{\boldsymbol H}_2\in{\mathfrak H}\, .
\ee
With these conditions, the flat connection \eqref{siqjkhdsjh2} becomes a quasiperiodic $1$-form:
\be
{\boldsymbol A}_\sigma(t,x+R)={\boldsymbol H}_2^{-1}\,{\boldsymbol A}_\sigma(t,x)\,{\boldsymbol H}_2\,.
\ee
Let us define the monodromy matrix at the time slice $t_0$  by
\bea\label{monoM1}
{\boldsymbol M}(\rho)=
\overset{\leftarrow}{{\cal P}}\exp\Big(\int_{0}^{R}{\rm d} x\  {\boldsymbol A}_x\Big)\Big|_{t=t_0}\
\ \ \ \ \ (\rho\equiv\rho_+)
\eea
with ${\boldsymbol A}_x={\boldsymbol A}_+-{\boldsymbol A}_-$.
Here the dependence on $\rho\equiv \rho_+$ is indicated explicitly though, of course, the monodromy matrix also depends on
$\varepsilon_{1}$, $\varepsilon_2$,
while $\rho_-$ is expressed in terms of these parameters using \eqref{hassyayt}. 
Then a textbook calculation shows that
\be
T(\rho)={\rm Tr}\big[{\boldsymbol H}_2{\boldsymbol M}(\rho)\big]
\ee
is independent of the choice of the time slice $t_0$ so that it can be thought 
of as the generating function of a continuous family of conserved charges.
In the contemporary paradigm of integrability in $1+1$ dimensional field theory
it is crucial to prove that these conserved charges mutually Poisson commute, i.e.,
\be\label{asdasdasd}
\big\{T(\rho_1),T(\rho_2)\big\}=0 
\ee
for arbitrary $\rho_1\ne\rho_2$. 
Owing to the complicated and non-ultralocal form of the Poisson brackets
$\big\{{\boldsymbol A}_x(x_1),\,{\boldsymbol A}_x(x_2)\big\}$,
the relations \eqref{asdasdasd} are far from evident (see e.g.\!\cite{Delduc:2015xdm}).
\bigskip

For $\varepsilon_1=\varepsilon_2=0$ (which corresponds to the PCF)
the computation of the Poisson brackets of the monodromy matrix was discussed in ref.\!\cite{Duncan:1989vg}.
In this case, the formula \eqref{currentsK} for the currents  becomes 
$
{\boldsymbol I}_\pm=
-2\ri\,
{\boldsymbol U}^{-1}\partial_\pm  {\boldsymbol U}
$. 
Assuming that $\rho_{\pm}=1-\varepsilon_2\,\zeta_\pm$ and
$\zeta_\pm$ are kept fixed as $\varepsilon_{1,2}\to 0$,
eq.\,\eqref{siqjkhdsjh2} turns into the Zakharov-Mikhailov connection \cite{Zakharov:1973pp}
\be\label{ZMconnection}
\lim_{\varepsilon_1,\varepsilon_2\to 0}{\boldsymbol A}_\pm=\,-\zeta_\pm^{-1}\ 
{\boldsymbol U}^{-1}\partial_\pm  {\boldsymbol U} \  ,
\ee
while the constraint \eqref{hassyayt} boils down to the relation $\zeta_++\zeta_-=2$.
The monodromy matrix for the PCF can be defined 
by taking the limit of \eqref{monoM1}:
\be
{\boldsymbol M}^{(0)}(\zeta)=\lim_{\varepsilon_1,\varepsilon_2\to 0}{\boldsymbol M}(\rho)\big|_{\rho=1-\varepsilon_2\zeta_+}\,,\ \ \
 \ {\rm where} \ \ \ \ \zeta_\pm\equiv 1\pm\zeta\, .
\ee
In ref.\!\cite{Duncan:1989vg}, for overcoming the non-ultralocality problem, the authors proposed a certain formal regularization procedure which 
results in the Yang-Baxter Poisson algebra
\be\label{owowowowow}
\big\{{\boldsymbol M}^{(0)}(\zeta_1) \begin{array}{ccc} \\[-0.4cm] \otimes \\[-0.35cm] , \end{array}{\boldsymbol M}^{(0)}(\zeta_2)\,\big\}=
\big[{\boldsymbol M}^{(0)}(\zeta_1)\,{\otimes}\, {\boldsymbol M}^{(0)}(\zeta_2),\, {\boldsymbol r}^{(0)}(\zeta_1-\zeta_2)\,\big]
\ee
with
\be
{\boldsymbol r}^{(0)}(\zeta_1-\zeta_2)=\ -\frac{{\rm g}^2}{2}\ \,\frac{q^{ab}\ {\tt t}_a\otimes{\tt t}_b}{\zeta_1-\zeta_2}\, . 
\ee
\bigskip

Of course, eq.\,\eqref{owowowowow} complemented by
$
\big[{\boldsymbol H}_2\otimes {\boldsymbol H}_2,\,{\boldsymbol r}^{(0)}(\zeta)\big]=0
$,
immediately implies the desired commutativity conditions  \eqref{asdasdasd} specialized to the 
PCF.  However, for the general Klim\v{c}\'{i}k model it is uncertain whether the classical Yang-Baxter Poisson algebra 
emerges, even at the formal level. Below we'll try to unravel this problem for 
$\mathfrak{G}=SU(2)$ by using results obtained in Section \ref{sec3}.
As before our considerations are inspired by the quantum case
and it will be useful to keep the following few aspects of the quantum model in mind.

\bigskip

\subsection{RG flow\label{sec43}}

Similar to the PCF, there is strong evidence
to suggest that the  integrability of the Klim\v{c}\'{i}k model extends to the quantum level. 
Among other things, this implies the perturbative renormalizability of the model.
In fact, one loop renormalizability was demonstrated for a more general class
of field theories in the work \cite{Valent:2009nv}. 
The RG flow equations describing the cutoff dependence  of the
bare coupling constants are given by \cite{Fateev:1996ea,Sfetsos:2015nya}
(see also Appendix \,\ref{AppB} for  some details)\footnote{Usually the 
Killing form in the definition of the Lagrangians \eqref{oaspsaop0},\,\eqref{oaspsaop} for a simple compact Lie group $\mathfrak{G}$ is understood
as a matrix trace over the fundamental irrep such that ${\rm Tr}({\tt t}_a{\tt t}_b)=\tfrac{1}{2}\delta_{ab}$.
This is related to our definition \eqref{iaosidosa}
as
$
\langle {\mathfrak a},{\mathfrak b} \rangle=\tfrac{1}{2}\,C_2(\mathfrak{G})\,{\rm Tr}(
{\mathfrak a}{\mathfrak b})$, where $C_2(\mathfrak{G})$ stands for the quadratic
Casimir in the adjoint representation. 
The advantage of our 
 convention is that the RG flow equations \eqref{hassasaty} do not involve any 
group dependent factors.}
\bea\label{hassasaty}
&&
\partial_\tau \varepsilon_1=-\tfrac{1}{2}\  \hbar \,{\rm g}^2\varepsilon_1 \ 
\big(1-(\varepsilon_1-\varepsilon_2)^2\big)\, \big(1-(\varepsilon_1+\varepsilon_2)^2\,\big)+O(\hbar^2)\nonumber\\[0.1cm]
&&\partial_\tau
(\varepsilon_2/\varepsilon_1)=O(\hbar^2)
\\[0.2cm]
&&\partial_\tau({\rm g}^2\varepsilon_{{ 1}})=O(\hbar^2) \nonumber
\eea
with 
$
\partial_\tau\equiv 2\pi\,  {\Lambda}\,{\textstyle \frac{{\textstyle \partial}}{{\textstyle \partial \Lambda}}}\,.
$
The second equation in \eqref{hassasaty} shows that
\be
\nu^2=\frac{\varepsilon_2}{\varepsilon_1}
\ee
is an RG invariant and the third equation is fulfilled  if we choose
\be\label{ioweiowe}
{\rm g}^2= \bigg|\frac{\varepsilon_1+\varepsilon_2}{\varepsilon_1\varepsilon_2}\bigg|\, .
\ee
This way in the quantum theory there is only one $\Lambda$-dependent bare coupling.
Within the domain 
$$0<\varepsilon_1,\varepsilon_2<1$$  
which will be considered in these notes,
it is convenient to use the parameterization
\be\label{kappdef}
\varepsilon_1=\frac{1}{\sqrt{(1+\kappa^{-1}\,\nu^2)(1+\kappa \nu^2)}}\,, \ \ \ \ \ \ 
\varepsilon_2=\frac{\nu^2}{\sqrt{(1+\kappa^{-1}\,\nu^2)(1+\kappa \nu^2)}}\
\ee
where $\nu^2>0$ and
\be
\kappa=\kappa(\Lambda)\ : \ \ \ \ 0<\kappa<1\,.
\ee
It follows from the RG flow equations \eqref{hassasaty} that a consistent removal of the UV cutoff $\Lambda$
requires that
\be\label{Llimitk}
\lim_{\Lambda\to\infty}\kappa(\Lambda)= 1^-\, .
\ee
Thus in the high energy limit the renormalized running coupling
will tend to one from below.  

\vskip 0.3in

\section{Monodromy matrix for the Fateev model}

\bigskip
Choosing a local co-ordinate frame  $\{X^\mu\}$
on  the group manifold  ${\mathfrak G}$, the  Klim\v{c}\'{i}k Lagrangian can be written in the form
\bea\label{aisiaosioa}
{\cal L}\,=\,2\,G_{\mu\nu}(X)\,\partial_+X^\mu\partial_-X^\nu-
B_{\mu\nu}(X)\,\big(\partial_+X^\mu\partial_-X^\nu-\partial_-X^\mu\partial_+X^\nu\,\big)\,.
\eea
Field theories of this type are known as non-linear sigma models and  describe the
propagation of a string  on a Riemannian manifold (the target space).
Interested readers can find some details concerning the target space background for the general model 
in Appendix\ \ref{AppB}.
Below we will focus on the simplest case with group ${\mathfrak G}=SU(2)$
where the target space is topologically equivalent to the three sphere.
With this choice, the $B$-term in \eqref{aisiaosioa} is a total derivative and can be ignored \cite{Hoare:2015gda}
and the theory coincides with the model originally introduced by Fateev in \cite{Fateev:1996ea}. The 
zero curvature representation for the Fateev model was found in \cite{Lukyanov:2012zt}
in a gauge which is different but equivalent to that of \eqref{siqjkhdsjh2} specialized to the case ${\mathfrak G}=SU(2)$
(the exact relation can be found in Appendix\ \ref{AppC}).
In both gauges, the Poisson brackets of the connection do not possess the ultralocal property and 
it is unknown whether an 
``ultralocal'' gauge actually exists  except for the cases
with $\varepsilon_2/\varepsilon_1=0,\infty$ considered in \cite{Bazhanov:2017nzh}. 
Thus, with a view towards first principles quantization, the Poisson algebra generated by the monodromy matrices
is of prime interest for the Fateev model and more generally the Klim\v{c}\'{i}k one.
\bigskip

In the context of quantization, the target space with
$\kappa\to 1^-$ deserves special study.
For this purpose, we introduce a co-ordinate frame
based on the  Euler decomposition for the group element
\bea\label{skdjfsd}
{\boldsymbol U}=\re^{-\frac{\ri v}{2}{\tt h}}\ \re^{-\frac{\ri\theta}{2}({\tt e}_++{\tt e}_-)}\ \re^{-\frac{\ri w}{2}{\tt h}}\, ,
\eea
where  ${\tt h},{\tt e}_\pm$ are the generators of the Lie algebra ${\mathfrak g}=\mathfrak{sl}_2$ \eqref{comm4}.
In fact, it is useful to substitute the angle $\theta\in(0,\pi)$ for $\phi\in(-\infty,\infty)$ such that
\bea\label{oqiwoeiw}
\tan(\tfrac{\theta}{2})=\ \ \re^{\phi-\phi_0}\,, \ \ \ \ \ \ 
 \re^{\phi_0}=\sqrt{\frac{1+\kappa}{1-\kappa}}\ \,.
\eea
In this frame, the symmetry
${\boldsymbol U}\mapsto {\boldsymbol H}_1{\boldsymbol U}{\boldsymbol H}_2$
(${\boldsymbol H}_1,{\boldsymbol H}_2\in\mathfrak{H}$) 
of the general Klim\v{c}\'{i}k model is manifested as the
invariance of the Fateev model
w.r.t. the  constant shifts
\be
v\mapsto v+v_0\,, \ \ \ \ \ \ \ \ w\mapsto w+w_0\, .
\ee
The corresponding Noether currents will be denoted by
$j^{(v)}$ and $j^{(w)}$ respectively.
With the continuity equations
\be
\partial_+j^{(A)}_-+\partial_-j^{(A)}_+=0\ \ \ \ \ \ \ \ (A=v,w)
\ee
one can introduce the dual fields $\tilde{v},\,\tilde{w}$ through the relations
\be
 j^{(v)}_\pm=\pm\,\partial_\pm\,\tilde{v}\,,\ \ \ \ \ \qquad
j^{(w)}_\pm=\pm\,\partial_\pm\,\tilde{w}\, \, .
\ee
It turns out that the dual field ${\boldsymbol \omega}$ defined 
by eq.\,\eqref{woeiowei} coincides with
\be
{\boldsymbol \omega}=\frac{1}{2}\,\bigg[\sqrt{1+\nu^2}\ \tilde{w}+\frac{\ri}{2}\log\bigg(\frac{\cosh(\phi_0+\phi)}{\cosh(\phi_0-\phi)}\bigg)\,\bigg]\,{\tt h}
\, .
\ee
The boundary conditions \eqref{asklas} specialized for the $SU(2)$ case with
\be
{\boldsymbol H}_{1}=\re^{-\ri\pi k_1{\tt h}}\,, \ \ \ \ \ {\boldsymbol H}_{2}=\re^{-\ri\pi k_2{\tt h}}\,,
\ee
imply the following conditions imposed on the fields $(\phi,v,w)$:
\be\label{boundary1}
 \phi(t,x+R)=\phi(t,x)\,,\ \ \ \ v(t,x+R)=v(t,x)+2\pi k_1\,, \ \ \ \ w(t,x+R)=w(t,x)+2\pi k_2 \,.
\ee
Also we will focus on the neutral sector of the model, which means periodic boundary conditions
for the dual fields
\be\label{boundary2}
{\tilde v}(t,x+R)={\tilde v}(t,x)\,, \ \ \ \ \ \ \ {\tilde w}(t,x+R)={\tilde w}(t,x)\,.
\ee

Taking into account that 
\be
{\hat {\boldsymbol R}}\big({\tt  h}\big)= 0\,,  \ \ \ \ \ \ \ {\hat {\boldsymbol R}}\big({\tt  e}_\pm\big)=\mp\ri\, {\tt  e}_\pm
\nonumber
\ee
and using  the parameterization \eqref{skdjfsd},\,\eqref{oqiwoeiw}
the Lagrangian \eqref{oaspsaop} with ${\rm g}^2$ as in \eqref{ioweiowe} 
can be expressed in terms  of three real fields
$(\phi,v,w)$ and two real parameters $\kappa$ and $\nu$ \eqref{kappdef}.
Here there is no need to present the explicit formula, 
we just note that for $|\phi|\ll \phi_0$ 
the Fateev Lagrangian 
can be approximated by  (up to a total derivative) 
\bea\label{LLag}
{\cal L}_{{\rm F}}
\asymp 2\,\Big(\partial_+\phi\,\partial_-\phi+\frac{1}{1+\nu^{-2}}\   \partial_+v\,\partial_-v+
\frac{1}{1+\nu^2}\  \partial_+w\,\partial_-w\ \Big)\ .
\eea
This implies that as $\kappa\to 1^-$, i.e., $\phi_0\to\infty$  
most of the target manifold asymptotically approaches the flat cylinder
with metric
$G_{\alpha\beta}\,{\rm d}X^\alpha{\rm d}X^\beta =({\rd}\phi)^2+(1+\nu^{-2})^{-1}({\rd}v)^2+(1+\nu^{2})^{-1}({\rd}w)^2$ while
the curvature is concentrated at the tips corresponding to $\phi=\pm\infty$. 
In the asymptotically flat domain, the general solution to the equations of motion 
can be expressed in terms of six arbitrary functions $\phi_i$ and $\bar{\phi}_i$:
\bea\label{fields11}
&&v(t,x)\asymp \sqrt{1+\nu^{-2}}\ \big(\phi_1(x_+)+\bar{\phi}_1(x_-)\big)\,, \ \ \ \ \ \ \
w(t,x)\asymp \sqrt{1+\nu^2}\ \big(\phi_2(x_+)+\bar{\phi}_2(x_-)\big) \nonumber \\[0.2cm]
&&\phi(t,x)\asymp\phi_3(x_+)+\bar{\phi}_3(x_-)\,,
\eea
while for the dual fields one has
\be\label{Dfields11}
{\tilde v}(t,x)\asymp \phi_1(x_+)-\bar{\phi}_1(x_-)\,, \ \ \ \ \ \ \
{\tilde w}(t,x)\asymp \phi_2(x_+)-\bar{\phi}_2(x_-) \, .
\ee

Having clarified the geometry of the target manifold for $\kappa\to1^-$
one can turn to the form of the flat connection \eqref{siqjkhdsjh2} in this limit.
We assume that the co-ordinates $(\phi,v,w)$ are kept within the 
asymptotic domain where eqs.\,\eqref{fields11},\,\eqref{Dfields11} are valid. 
Also, since the product  $\rho_+\rho_-$ \eqref{hassyayt} vanishes as
$1-\kappa$, we keep $\rho_+$ fixed while 
$\rho_-\to 0$. 
Then a direct calculation shows that 
\be\label{ksdjksd}
\lim_{\kappa\to 1^-\atop
\rho_+ - {\rm fixed}
} \Big(\partial_+- (\rho_+/\rho_-)^{+\frac{{\tt h}}{4}}\
{\boldsymbol A}_+^{({ \omega})}\,  
(\rho_+/\rho_-)^{-\frac{{\tt h}}{4}}\,\Big)=
\re^{+2\ri{\boldsymbol \omega}_+(x_+)}\,\big(\partial_+-{\boldsymbol B}(x_+|\rho_+)\,\big)\,\re^{-2\ri{\boldsymbol\omega}_+(x_+)}\,,
\ee
where we have used the gauge ${\boldsymbol A}_+^{({ \omega})}$ from eq.\,\eqref{oiqwoieoiw}. The 1-form
${\boldsymbol B}$ in this equation is defined by \eqref{Atilde1},\,\eqref{cvertex1},\,\eqref{cvertex0} and
\bea
{\boldsymbol \omega}_+(x_+)=\tfrac{1}{2}\,\big(\sqrt{1+\nu^2}\ \phi_2(x_+)+\ri\,\phi_3(x_+)
                                                             \big)\,{\tt h}\,.
\eea
For the other connection component one finds
\be\label{ksdjksd2}
\lim_{\kappa\to 1^-\atop
\rho_+ - {\rm fixed}}
 (\rho_+/\rho_-)^{+\frac{{\tt h}}{4}}\ {\boldsymbol A}_-^{(\omega)}\, 
 (\rho_+/\rho_-)^{-\frac{{\tt h}}{4}}=0\,.
\ee

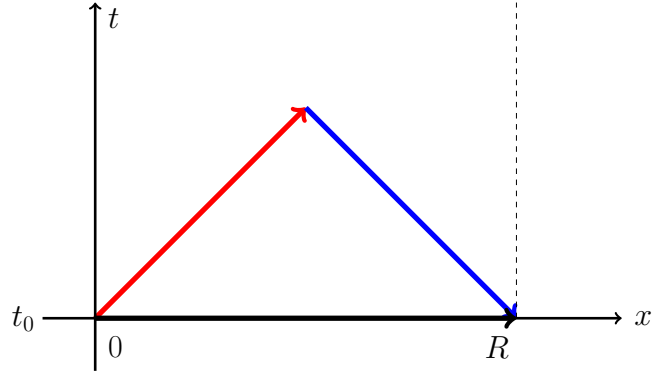
\begin{figure}[t]
\centering
\scalebox{0.7}{
\begin{tikzpicture}
\draw [line width = 1mm,->,red] (0,0) -> (4,4);
\draw [line width = 1mm,->,blue] (4,4) -> (8,0);
\draw [ line width = 1mm,->] (0,0) -> (8,0);
\draw [line width = 0.5mm, ->] (0,-1) -- (0,6);
\draw [line width = 0.5mm,->] (-1,0)-- (10,0);
\draw [dashed] (8,0) -- (8,6);
\node [below right] at (0.1,-0.2) {\Large $0$};
\node [below left] at (8,-0.2) {\Large $R$};
\node [left] at (-1,0) {\Large $t_0$};
\node [right] at (0.1,5.7) {\Large $t$};
\node [right] at (10.1,0) {\Large $x$};
\end{tikzpicture}
}
\caption{The integration along the time slice $t\,=\,t_0$ (black arrow) in eq.\,\eqref{monoM2}
 can be replaced by an integration 
along the characteristics: $x_-\,=\,t_0$ with 
$t_0<x_+<t_0+R$ (red arrow) and $x_+\,=\,t_0+R$ with $t_0<x_-<t_0-R$ (blue arrow).}
\label{fig03}
\end{figure}
We now turn to 
the monodromy matrix that was  introduced previously in \eqref{monoM1}.
In light of eqs.\,\eqref{ksdjksd},\,\eqref{ksdjksd2} we express
${\boldsymbol M}(\rho)$ in terms of ${\boldsymbol A}^{(\omega)}_\sigma$:
\bea\label{monoM2}
{\boldsymbol M}(\rho)=
\re^{-\ri{\boldsymbol \omega}(t_0,R)}\,\overset{\leftarrow}{{\cal P}}\exp\Big(\int_{0}^{R}{\rm d} x\  {\boldsymbol A}_x^{(\omega)}\Big)\Big|_{t=t_0}\ \re^{\ri{\boldsymbol \omega}(t_0,0)}
\ \ \ \ \ (\rho\equiv\rho_+)\,.
\eea
Since the connection ${\boldsymbol A}_\sigma^{(\omega)}$ is flat, the integral  over the segment $(0,R)$
can be transformed into the piecewise integral over the light cone segments as shown in fig.\,\ref{fig03}.
The monodromy matrix is then expressed
in terms of the light cone values of the connection as
\be
{\boldsymbol{ M}}(\rho)=\re^{-\ri{\boldsymbol \omega}(t_0,R)}\
{\overset{\leftarrow} {\cal P}}\exp\bigg(\int_{t_0}^{t_0-R}%
{\boldsymbol A}_-^{(\omega)}(x_-)\, \rd x_- \bigg)\,
{\overset{\leftarrow} {\cal P}}
\exp\bigg(\int_{t_0}^{t_0+R} {\boldsymbol A}_+^{(\omega)}(x_+)\,\rd x_+\bigg)\ \re^{\ri{\boldsymbol \omega}(t_0,0)}
\ee
where
\be
{\boldsymbol A}_+^{(\omega)}(x_+)={\boldsymbol A}_+^{(\omega)}(t,x)\big|_{x_-=t_0}\,, \ \ \ \ \ \ \ 
{\boldsymbol A}_-^{(\omega)}(x_-)={\boldsymbol A}_-^{(\omega)}(t,x)\big|_{x_+=t_0+R}\ .
\ee
For $\kappa$ close to $1$ 
the instant $t_0$ can be chosen such that the values of the fields
lie in the asymptotically flat region of the target manifold where 
formulae \eqref{fields11},\,\eqref{Dfields11} are applicable. 
Then with eqs.\,\eqref{ksdjksd},\,\eqref{ksdjksd2}
at hand,
it is straightforward to show that the following limit exists
\be\label{redefine}
\lim\limits_{\kappa\to 1^-\atop \rho_+ - {\rm fixed}}(\rho_+/\rho_-)^{+\frac{{\tt h}}{4}}
\,{\boldsymbol{ M}}(\rho)\,
(\rho_+/\rho_-)^{-\frac{{\tt h}}{4}}
={\boldsymbol{ M}}^{(1)}(\rho)\,.
\ee
Explicitly, ${\boldsymbol{ M}}^{(1)}(\rho)$  can be expressed in terms of ${\boldsymbol L}_{\rm cl}(\rho)$
previously defined in \eqref{Lop3} and \eqref{Atilde1}:
\be
{\boldsymbol{ M}}^{(1)}(\rho)=
{\boldsymbol \Omega}^{-1}\
{\boldsymbol L}_{\rm cl}(\rho)\, \re^{\pi(2\ri\sqrt{1+\nu^2}P_2-P_3)\,{\tt h}}
\ {\boldsymbol \Omega}\,.
\ee
Here we take into account that $\phi(t_0,x+R)=\phi(t_0,x)$, ${\tilde w}(t,x+R)={\tilde w}(t,x)$ and use
\bea\label{Pdef}
P_3&\equiv&\tfrac{1}{2\pi}\big(\phi_3(t_0+R)-\phi_3(t_0)\,\big)=-\tfrac{1}{2\pi}\big(\bar{\phi}_3(t_0-R)-\bar{\phi}_3(t_0)\,\big)\\[0.2cm]
P_2&\equiv&\tfrac{1}{2\pi}\big(\phi_2(t_0+R)-\phi_2(t_0)\,\big)=+\tfrac{1}{2\pi}\big(\bar{\phi}_2(t_0-R)-\bar{\phi}_2(t_0)\,\big) \nonumber
\eea
  and
\be\label{omega0def}
{\boldsymbol \Omega}=\re^{-\frac{\ri}{2}{ \omega}_0{\tt h}}\ : \ \ \ \ \ 
{\omega_0}=w(t_0,R)+\ri\,\big(\phi_3(t_0+R)-\bar{\phi}_3(t_0-R)\big)\ .
\ee

It follows from the Lagrangian that the chiral fields $\phi_i$ 
can be chosen to satisfy the Poisson bracket relations 
\be\label{pqwoqpwo}
\{\phi_i(x_+),\,\phi_j(x_+')\}=-\tfrac{1}{4}\,\delta_{ij}\,\epsilon(x_+-x_+')
\ee
and hence, using the results of the previous section,
${\boldsymbol L}_{\rm cl}(\rho)$ obeys the Yang-Baxter Poisson algebra
\eqref{PBrel11111}. 
In the Hamiltonian picture the boundary condition $w(t,x+R)=w(t,x)+2\pi k_2$
with $k_2$ a non-dynamical constant
is a constraint of the first kind $\grave{{\rm a}}$ la Dirac which
 should be supplemented by a gauge fixing condition.
Considering the fields in the asymptotically flat domain
where formulae \eqref{fields11},\,\eqref{Dfields11} hold true
leads to the relation
\be
P_2=\frac{k_2}{2\sqrt{1+\nu^2}}
\ee
and the gauge fixing condition can be chosen as
$
w(t_0,R)=0\,.
$
This way $\omega_0$ in \eqref{omega0def} becomes 
$
\omega_0=\ri\,\big(\phi_3(t_0+R)-\bar{\phi}_3(t_0-R)\big).
$
Similarly, we supplement the periodic boundary condition $\phi(t_0,x+R)=\phi(t_0,x)$ by the constraint
$\bar{\phi}_3(t_0-R)=0$,
so that 
\be\label{omega02}
\omega_0=\ri\,\phi_3(t_0+R)\,.
\ee
The Poisson brackets of ${\boldsymbol{ M}}^{(1)}(\rho)=
{\boldsymbol \Omega}^{-1}\
{\boldsymbol L}_{\rm cl}(\rho)\ \re^{\pi(\ri  k_2-P_3)\,{\tt h}}
\ {\boldsymbol \Omega}$ are obtained
by using \eqref{PBrel11111} and
the simple relations
\be\arraycolsep=0.3cm
\begin{array}{lll}
\big\{{\boldsymbol L}_{\rm cl}(\rho),\pi P_3\big\}=\tfrac{1}{4}\,\big[\,{\tt h},{\boldsymbol L}_{\rm cl}(\rho)\,\big]\,, 
 & \big\{{\boldsymbol L}_{\rm cl}(\rho),\omega_0\big\}
=\tfrac{\ri}{4}\,{\tt h}\, {\boldsymbol L}_{\rm cl}(\rho)\,,
& \big\{\omega_0,\pi P_3\big\}=\tfrac{\ri}{4}\,.
\end{array}
\ee
The latter follow from eqs.\,\eqref{Pdef},\,\eqref{pqwoqpwo},\,\eqref{omega02}.
Also, taking into account that
\be
\big[1\otimes {\tt h}+{\tt h}\otimes 1, {\boldsymbol r}(\lambda)\big]=0\,,
\ee
one arrives at
\be\label{PBrel22}
\big\{{\boldsymbol{ M}}^{(1)}(\rho_1)\begin{array}{ccc} \\[-0.4cm] \otimes \\[-0.35cm] , \end{array}{\boldsymbol{ M}}^{(1)}(\rho_2)\,\big\}=
\big[{\boldsymbol{ M}}^{(1)}(\rho_1)\,{\otimes}\, {\boldsymbol{ M}}^{(1)}(\rho_2), {\boldsymbol r}
(\lambda_1/\lambda_2)\,\big]\ ,
\ee
where recall that  $\rho_{1,2}$  depend on $\lambda_{1,2}$ via the relation \eqref{func1}.
\bigskip

It should be highlighted that the Poisson algebra \eqref{PBrel22} was obtained for a certain choice of the time slice $t_0$ 
when the fields take values in the asymptotic region. 
The validity of this equation for an arbitrary choice of $t_0$ is debatable, since the monodromy matrix
itself is not a conserved quantity. However that eq.\,\eqref{PBrel22} holds true 
even for a particular value of $t_0$ is sufficient
to prove 
the commutativity condition $
\{T^{(1)}(\rho_1),T^{(1)}(\rho_2)\}=0 
$ with 
\be
T^{(1)}(\rho)={\rm Tr}\big[\re^{-\ri\pi k_2{\tt h}}{\boldsymbol{ M}}^{(1)}(\rho)\big]=
\lim\limits_{\kappa\to 1^-\atop \rho_+ - {\rm fixed}}{\rm Tr}\big[\,\re^{-\ri\pi k_2{\tt h}}
\,{\boldsymbol{ M}}(\rho)
\big]\,.
\ee

In view of the above, it makes sense to
 reconsider our definition of the monodromy matrix for the Fateev model and
introduce
\be\label{oqoqowiwieu}
{\boldsymbol M}^{(\kappa)}(\rho)=
(\rho_+/\rho_-)^{+\frac{{\tt h}}{4}}
\,{\boldsymbol{ M}}(\rho)\,
(\rho_+/\rho_-)^{-\frac{{\tt h}}{4}} \ \ \ \ \ \ \ (\rho\equiv\rho_+)\, .
\ee
We've just seen that in the $\kappa\to 1^-$ limit, the matrix ${\boldsymbol M}^{(\kappa)}(\rho)$ 
obeys the Yang-Baxter Poisson algebra \eqref{PBrel22}. On the other hand,
the redefinition \eqref{oqoqowiwieu} has no effect on the monodromy matrix as $\kappa\to 0$ and
 both $\rho_\pm\to 1$ so that the Yang-Baxter algebra is still satisfied but in the form 
\eqref{owowowowow}. Finally, the case $\nu=0$ with $\kappa\in(0,1)$ 
was already considered in the work \cite{Bazhanov:2017nzh} where it was shown that
\be\label{PBrel33}
\big\{{\boldsymbol{ M}}^{(\kappa)}(\rho_1)\begin{array}{ccc} \\[-0.4cm] \otimes \\[-0.35cm] , \end{array}{\boldsymbol{ M}}^{(\kappa)}(\rho_2)\,\big\}=
\big[{\boldsymbol{ M}}^{(\kappa)}(\rho_1)\,{\otimes}\, {\boldsymbol{ M}}^{(\kappa)}(\rho_2), {\boldsymbol r}
(\lambda_1/\lambda_2)\,\big]\ \ \ \ \ \ \ (\nu\to 0)
\ee
with $\rho_{1,2}=\lambda_{1,2}$. All this suggests that the key relations \eqref{PBrel33} may extend to
the parametric domain $\nu^2> 0$ and $\kappa\in(0,1)$   with some function $\rho=\rho(\lambda|\nu,\kappa)$
(which is unknown in general).

\section{Conclusion}

For classically integrable field theories, the Yang-Baxter Poisson algebra 
plays a r$\hat{{\rm o}}$le similar to that of the canonical Poisson bracket 
relations for a general mechanical system. 
Whereas the correspondence principle 
prescribes the replacement of the canonical Poisson brackets
with commutators,
the ``first principles'' quantization in integrable models 
starts with the formal substitution of the Yang-Baxter Poisson algebra by 
the quantum Yang-Baxter algebra.
However, many interesting models possessing the zero curvature representation
belong to the  non-ultralocal  class of theories where it is
 difficult to ascertain
 the emergence of the Yang-Baxter Poisson algebra.
This makes the quantization of such models problematic.

In this work, we investigated the emergence of the Yang Baxter Poisson algebra
in a non-ultralocal system. Our considerations are inspired by the age-old observation
that the quantum monodromy operator is somehow better 
behaved than its classical counterpart. 
In our central example we recovered
the Yang-Baxter Poisson
algebra in a non-ultralocal system based on the $SU(2)$ current algebra  by
starting with an explicit quantum field theory realization of the 
Yang-Baxter relation and then taking the classical limit.
As a result of the  entangled interplay between the 
classical limit and the
scaling one, which required ultraviolet regularization
of the model,
we found that the classical monodromy matrix
is somewhat more cumbersome than its quantum counterpart.
It turned out that the net result of the non-ultralocal structure 
for the Yang-Baxter Poisson algebra is the non-universal 
renormalization of the spectral parameter which occurs even 
at the classical level. This is somewhat in the spirit of 
Faddeev and Reshetikhin \cite{Faddeev:1985qu}
who proposed to ignore the problem of non-ultralocality,
arguing that it is a consequence of 
choosing the ``false vacuum'',  and to restore the ultralocality of the 
current algebra by hand. 

The example we elaborated is relevant
to the Fateev model, an integrable two parameter 
deformation of the $SU(2)$ Principal Chiral Field.
It provides evidence for the existence of the
Yang-Baxter Poisson structure for this
remarkable non-linear sigma model,
which was shown for several particular cases in the parameter space.
We believe that 
unraveling the Yang-Baxter Poisson algebra
for non-ultralocal systems
is important in many respects.
Of special interest 
is the Klim\v{c}\'{i}k model and its reductions \cite{Delduc:2013fga}
which have recently attracted a great deal of attention
in the context of the AdS/CFT correspondence \cite{Delduc:2013qra,Arutyunov:2013ega}.
We supplement these notes by two appendices 
which collect a number of facts about the 
Klim\v{c}\'{i}k model
that, in our opinion, fill some gaps in the current literature.

\bigskip

\noindent
{\bf Note added.}  
In the previous version in Appendix A, a  formula was presented relating the currents
${\bm I}_\sigma$ and ${\bm J}_\sigma$. It turns out to admit a significant simplification.
This has allowed us to shorten the presentation 
by transferring the simplified formula to the main body of the text,
see eq.\,\eqref{aioasaso}, and removing Appendix A entirely. 
Parts of  what used to be Appendix A have been incorporated into
sec.\ref{sec4} and some accompanying 
minor stylistic changes were made, e.g., the splitting of sec.\ref{sec4}
into subsections and the removal of some redundant formulae.

\section{Acknowledgments}
Part of the work was done during the third author's visit to  the International Institute of Physics at Natal 
and KITP at Santa Barbara in the fall of 2017. S.L. would like to thank these institutes for the support and
hospitality he received during his stay. The authors are grateful to Vladimir Mangazeev for
useful suggestions as well as his interest in this work.

 \appendix

\section{Appendix\label{AppB}}
Here we discuss some geometrical aspects of the Klim\v{c}\'{i}k non-linear sigma model.
The target space is topologically the same  as   ${\mathfrak G}$ (which below 
is assumed to be a compact simple Lie group)
but equipped with a certain 
 anisotropic metric  $G_{\mu\nu}$. The latter  can be thought of as a two-parameter deformation
of the left/right invariant metric on the group manifold.  In fact, the form of the Lagrangian   \eqref{aisiaosioa}
suggests  that the target manifold is
 equipped with  the  affine connection ${\mathsf\Gamma}$
such that the metric  
 is covariantly
constant w.r.t.  ${\mathsf\Gamma}$, while  its   torsion is  defined by the antisymmetric tensor $B_{\mu\nu}$.
To be precise,  the covariant torsion tensor
 \bea\label{torsion1}
H_{\lambda\mu\nu}\,=\,
G_{\lambda\rho}\ \big(
{{\mathsf \Gamma}^{\rho}}_{\mu\nu}-{{\mathsf \Gamma}^{\rho}}_{\nu\mu}\big)
\eea
(here ${{\mathsf\Gamma}^{\rho}}_{\mu\nu}$ stands for the Christoffel symbol),
is a closed 3-form with  $B_{\mu\nu}$ playing  the r${\hat {\rm o}}$le  of the torsion potential:
\bea
H_{\lambda\mu\nu}=\partial_\lambda B_{\mu\nu}+\partial_\nu B_{\lambda \mu}+
\partial_\mu B_{\nu\lambda}\ .\nonumber
\eea

A remarkable feature of the  Klim\v{c}\'{i}k target space background  is that it admits  a set of
  1-forms 
which can be thought of as deformations of the Maurer-Cartan forms.
Introduce   two sets   $\{{\mathlarger{\mathlarger{\mathpzc e}}}_\mu^a(\sigma)\}_{a=1}^D\ 
(D={\rm dim}\,\mathfrak{G})$: 
\bea\label{ajsaisui}
{\tt t}_a\, {\mathlarger{\mathlarger{\mathpzc e}}}_\mu^a(\sigma)\,\rd X^\mu\,=\,
-2\,\ri\ \hat{{\boldsymbol  \Omega}}^{-1}_\sigma\big(
{\boldsymbol U}^{-1}\,{\rm d}\,{\boldsymbol U}\big)\ .
\eea
Here 
$\hat{{\boldsymbol  \Omega}}_\sigma$   stands for the linear operator  acting in $\mathfrak{g}$,
\bea\label{defomega1}
\hat{{\boldsymbol  \Omega}}_\sigma\,=\,\hat{\boldsymbol 1}+\ri\,\sigma\,\varepsilon_1\,
\hat{\boldsymbol R}_{\boldsymbol U}+
\ri\,\sigma\,\varepsilon_2\,\hat{\boldsymbol R}
\eea
and 
$\sigma$ takes  two values $\pm\,$.
It is not difficult to show that  the metric can be written as
\bea\label{ajssuaaus}
&G_{\mu\nu}\,=\,\frac{1}{2{\rm g}^2}\,\,q_{ab}\ {\mathlarger{\mathlarger{\mathpzc e}}}^{a}_\mu(+) \,{\mathlarger{\mathlarger{\mathpzc e}}}^{b}_\nu(+)\,=\,
\frac{1}{2{\rm g}^2}\,\,q_{ab}\ {\mathlarger{\mathlarger{\mathpzc e}}}^{a}_\mu(-) 
\,{\mathlarger{\mathlarger{\mathpzc e}}}^{b}_\nu(-)\ ,
\eea
i.e., $\{{\mathlarger{\mathlarger{\mathpzc e}}}_\mu^a(+)\}_{a=1}^D$ and 
$\{{\mathlarger{\mathlarger{\mathpzc e}}}_\mu^a(-)\}_{a=1}^D$ are two   vielbein sets in the cotangent space  of the
target manifold.
Notice   the  following simple  relations
\bea
&G^{\mu\nu}\ {\mathlarger{\mathlarger{\mathpzc e}}}^a_\mu(+)\,{\mathlarger{\mathlarger{\mathpzc e}}}^b_\nu(+)\,=\,G^{\mu\nu}\ 
{\mathlarger{\mathlarger{\mathpzc e}}}^a_\mu(-)\, {\mathlarger{\mathlarger{\mathpzc e}}}^b_\nu(-)\,=\,2{\rm g}^2\ q^{ab}\nonumber
\eea
and 
\bea\label{jausauassa}
\sqrt{\det G^{\phantom{{(0)}}}_{\mu\nu}}\,=\,\big(\det\hat{{\boldsymbol \Omega}}_{\sigma}
\,\big)^{-1}\,\times\,\sqrt{\det {G^{{ (0)}}_{\mu\nu}}}\ ,
\eea
where  $  G^{(0)}_{\mu\nu}= G_{\mu\nu}|_{\varepsilon_1=\varepsilon_2=0}$\,.

It turns out that  the torsion also admits  simple expressions involving    ${\mathlarger{\mathlarger{\mathpzc e}}}_\mu^a(\sigma)$  and 
the structure constants  $ F^{abc} (\sigma,\sigma'|\sigma'')$ \eqref{osasoasa} appearing in the Poisson algebra \eqref{ajsasysya}:
\begin{subequations}\label{aasiais}
\begin{equation}
H_{\lambda\mu\nu}=+\frac{1}{4{\rm g}^2}\,
\big(F_{abc}(-\,+|+)\,{\mathlarger{\mathlarger{\mathpzc e}}}^{c}_{\,[\lambda}(+)
{\mathlarger{\mathlarger{\mathpzc e}}}^a_\mu(-){\mathlarger{\mathlarger{\mathpzc e}}}^b_{\nu]}(+)
-2\,F_{abc}(+\,+|+)\,{\mathlarger{\mathlarger{\mathpzc e}}}^a_{\lambda}(+)
 {\mathlarger{\mathlarger{\mathpzc e}}}^b_{\mu}(+){\mathlarger{\mathlarger{\mathpzc e}}}^c_{\nu}(+)\big)
\end{equation}
and
\begin{equation}
H_{\lambda\mu\nu}=-\frac{1}{4{\rm g}^2}\,\big(F_{abc}(+\,-|-)\,{\mathlarger{\mathlarger{\mathpzc e}}}^{c}_{[\lambda}(-)
{\mathlarger{\mathlarger{\mathpzc e}}}^a_\mu(+){\mathlarger{\mathlarger{\mathpzc e}}}^b_{\nu]}(-)
-2\,F_{abc}(-\,-|-)\,{\mathlarger{\mathlarger{\mathpzc e}}}^a_{\lambda}(-){\mathlarger{\mathlarger{\mathpzc e}}}^b_{\mu}
(-){\mathlarger{\mathlarger{\mathpzc e}}}^c_{\nu}(-)\big)\,.
\end{equation}
\end{subequations}
Here the symbol $[\lambda\mu\nu]$ denotes the alternating summation over all possible  permutations  of 
the indices  $\lambda,\mu$ and $\nu$.

\bigskip
Before discussing the origin of the above formulae for the metric and torsion,
let us first inspect the  reality condition for the target space background.
Consider the metric and the torsion as a function of $\varepsilon_1$ with 
the ratio $\varepsilon_2/\varepsilon_1$
a fixed { real} number.
First of all it is easy to see that the determinant  $\det\hat{{\boldsymbol \Omega}}_\sigma$  which appears in 
the formula  \eqref{jausauassa}
does not depend on the choice of the sign factor $\sigma$ -- it 
is a polynomial in the variable $\varepsilon_1^2$ of degree  coinciding with the  integer part
of  half of  $D\equiv{\rm dim} ({\mathfrak G})$:
\bea
\det\hat{{\boldsymbol \Omega}}_\sigma=1+\sum_{n=1}^{[\frac{D}{2}]} \omega^{(n)}\ \varepsilon_1^{2n}\ ,\nonumber
\eea
where the coefficients $ \omega^{(n)}$ are real as $\Im m(\varepsilon_2/\varepsilon_1)=0$.
In their  turn, the  components of the metric tensor and the torsion  are rational functions of $\varepsilon_1$ of the form
\bea\label{oaisoas}
G_{\mu\nu\phantom{\lambda}}&=&
\frac{1}{\det\hat{{\boldsymbol \Omega}}_\sigma}\ \sum_{n=0}^{[\frac{D-1}{2}]} g^{(n)}_{\mu\nu}\ \varepsilon_1^{2n}
\\[0.2cm]
H_{\lambda\mu\nu}&=&\frac{\ri\varepsilon_1}{(\det\hat{{\boldsymbol \Omega}}_\sigma)^2}\ 
\ \sum_{n=0}^{D-1}
h^{(n)}_{\lambda\mu\nu}\ \varepsilon_1^{2n}\ .\nonumber
\eea
For pure imaginary $\varepsilon_1$, the 1-forms ${\mathlarger{\mathlarger{\mathpzc e}}}_\mu^a(\sigma)$ 
are real and, as it follows from \eqref{ajssuaaus},
the metric  is positive definite. Formula \eqref{oaisoas} implies that it remains 
positive definite 
for sufficiently small real $\varepsilon_1$.\footnote{
Presumably the metric remains  positive  definite  in  the parameter  domain 
$ 0<\varepsilon_1< 1\ ,\   0<\varepsilon_2< 1-\varepsilon_1$.}
At the same time, as it follows from  \eqref{aasiais},\,\eqref{osasoasa} the torsion is real for pure imaginary $\varepsilon_1$. 
Therefore  the expansion coefficients $h^{(n)}_{\lambda\mu\nu}$ turn out to be real as $\Im m(\varepsilon_2/\varepsilon_1)=0$.
However,
 $H_{\lambda\mu\nu}$ takes  { pure imaginary values}
 for real $\varepsilon_1$ and $\varepsilon_2$, in particular 
 for $0<\varepsilon_1< 1$,   $0<\varepsilon_2< 1-\varepsilon_1$. Notice that
 the case ${\mathfrak G}=SU(2)$ turns out to be somewhat special in that the torsion becomes zero  identically \cite{Hoare:2015gda}.
 The corresponding non-linear sigma model
 is equivalent to the model introduced by Fateev in ref.\!\cite{Fateev:1996ea}.
 In the presence of non-vanishing torsion, the  Lagrangian \eqref{aisiaosioa} is not invariant 
 under the substitution $(t\pm x)\mapsto (t\mp x)$, i.e., the field theory is not  $P$-invariant.
However it is still  invariant w.r.t. the special 
Lorentz transformation $(t\pm x)\mapsto  \re^{\pm \theta}\, (t\pm x)$ with real $\theta$.

\subsubsection*{Vielbeins}

To clarify the special r$\hat{\rm o}$le  of  the 1-forms \eqref{ajsaisui} for the Klim\v{c}\'{i}k target space background
let us make the following observations.

\bigskip
First we  point out     that the 1-forms ${\mathlarger{\mathlarger{\mathpzc e}}}_\mu^a(+)$
are covariantly constant w.r.t. the spin-connection
\bea\label{gtrw}
{\omega_{\nu, a}}^b(+)\,=\,{{ F}_{ac}}{}^b(+\,-| +)\,{\mathlarger{\mathlarger{\mathpzc e}}}^{c}_\nu(-)\, ,\nonumber
\eea
i.e.,
\bea\label{COVCONST1}
\partial^{\vphantom a}_\nu\, {\mathlarger{\mathlarger{\mathpzc e}}}^{a}_\mu(+)\,-\,{
{\mathsf\Gamma}^\lambda}_{\mu\nu}\,{\mathlarger{\mathlarger{\mathpzc e}}}^{a}_\lambda(+)+{\omega_{\nu, b}}^a(+)\,{\mathlarger{\mathlarger{\mathpzc e}}}^{b}_\mu(+)=0\ .
\eea
A  simple consequence of this fact is that the covariant derivative of the metric   \eqref{ajssuaaus} is zero,
as it should be. In a similar manner,  the 1-forms   ${\mathlarger{\mathlarger{\mathpzc e}}}_\mu^a(-)$ satisfy  the  covariant constant condition
\bea\label{COVCONST2}
\partial^{\vphantom a}_\nu\,{\mathlarger{\mathlarger{\mathpzc e}}}^{a}_\mu(-)\,-\,
{{\mathsf\Gamma}^\lambda}_{\nu\mu}\,{\mathlarger{\mathlarger{\mathpzc e}}}^{a}_\lambda(-)+{\omega_{\nu, b}}^a(-)\, {\mathlarger{\mathlarger{\mathpzc e}}}^{b}_\mu(-)=0
\eea
which involves another spin-connection 
\bea\label{gtrwa}
{\omega_{\nu, a}}^b(-)\,=\, {{ F}_{ac}}{}^b(-\,+|-)\, 
{\mathlarger{\mathlarger{\mathpzc e}}}^{+}_\nu(+)\, .\nonumber
\eea
Finally, the  covariantly constant 1-forms obey  the  
Maurer-Cartan type
equations:
\bea\label{shhssy}
&\hspace{-1cm}\partial\vphantom{{\mathlarger{\mathlarger{\mathpzc e}}}}_{[\nu}^{\vphantom{a}}{\mathlarger{\mathlarger{\mathpzc e}}}_{\mu]}^{a}(+)-\frac{1}{2}\ \Big(
q^{aa'}\,F_{a'bc}(++|+)
-
 \Theta^{aa'}\, F_{a'bc}(-+|+)\,\Big)\,  {\mathlarger{\mathlarger{\mathpzc e}}}^b_{[\nu}(+)\, 
{\mathlarger{\mathlarger{\mathpzc e}}}^c_{\mu]}(+) =0\nonumber\\[-0.15cm]
&\\[-0.15cm]
&\hspace{-1cm}\partial\vphantom{{\mathlarger{\mathlarger{\mathpzc e}}}}_{[\nu}^{\vphantom{a}}{\mathlarger{\mathlarger{\mathpzc e}}}^{a}_{\mu]}(-)-\frac{1}{2}\ \Big(
q^{aa'}\,F_{a'bc}(--|-)
-\Theta^{a'a}\, F_{a'bc}(+-|-)\,\Big)\, {\mathlarger{\mathlarger{\mathpzc e}}}^b_{[\nu}(-)\,{\mathlarger{\mathlarger{\mathpzc e}}}^c_{\mu]}(-) =0\nonumber
\eea
with
\be\label{osssai}
\Theta^{aa'}\, :\ \ \ {\mathlarger{\mathlarger{\mathpzc e}}}^a_{\mu}(+)={\Theta^{a}}_b\ {\mathlarger{\mathlarger{\mathpzc e}}}^b_{\mu}(-)\ ,
\qquad
\Theta^{aa'}\, =\, {\textstyle \frac{1}{2{\rm g}^2}}\ G^{\mu\nu}\ {\mathlarger{\mathlarger{\mathpzc e}}}^a_{\mu}(+)\, 
{\mathlarger{\mathlarger{\mathpzc e}}}^{a'}_{\nu}(-)\,,
\qquad
{\Theta^{a}}_{c}\,q^{cd}\, {\Theta^{b}}_{d}=q^{ab}\,.\nonumber
\ee

\bigskip
Relations \eqref{COVCONST1},\,\eqref{COVCONST2} allow one to express the torsion in terms of
 ${\mathlarger{\mathlarger{\mathpzc e}}}^a_\mu(\sigma)$. Namely, a simple calculation yields
\bea\label{Connection1}
&{\mathsf\Gamma}_{\lambda\mu\nu}=\frac{1}{2{\rm g}^2}\,q_{ab}\ 
\Big(\,{\omega_{\nu,c}}^a(+)\,
{\mathlarger{\mathlarger{\mathpzc e}}}^b_\lambda(+)\,{\mathlarger{\mathlarger{\mathpzc e}}}^c_{\mu}(+)\,+\,{\mathlarger{\mathlarger{\mathpzc e}}}_\lambda^a(+)
\,\partial_\nu\,{\mathlarger{\mathlarger{\mathpzc e}}}_\mu^b(+)
\Big)\phantom{\,.}\nonumber\\[-0.2cm]
& \\[-0.2cm]
&{\mathsf\Gamma}_{\lambda\mu\nu}=\frac{1}{2{\rm g}^2}\,q_{ab}\ 
\Big(\,{\omega_{\mu, c}}^a(-)\,{\mathlarger{\mathlarger{\mathpzc e}}}^b_\lambda(-)\,
{\mathlarger{\mathlarger{\mathpzc e}}}^c_{\nu}(-)\,+\,{\mathlarger{\mathlarger{\mathpzc e}}}_\lambda^a(-)\,\partial_\mu\,{\mathlarger{\mathlarger{\mathpzc e}}}_\nu^b(-)
\Big)\,.\nonumber
\eea
These formulae, combined with \eqref{torsion1} imply
\begin{eqnarray}
\!\!\!\!\!\!\!\!\!\!&H_{\lambda\mu\nu}\,=\,\frac{1}{2{\rm g}^2}\,\sigma\, q_{ab}\
\Big(\,{\mathlarger{\mathlarger{\mathpzc e}}}^a_\lambda(\sigma)\,\big({\omega_{\nu, c}}^b(\sigma)\,{\mathlarger{\mathlarger{\mathpzc e}}}^c_{\mu}(\sigma)-{\omega_{\mu, c}}^b(\sigma)\,{\mathlarger{\mathlarger{\mathpzc e}}}^c_{\nu}(\sigma)
\big)\,+\,{\mathlarger{\mathlarger{\mathpzc e}}}_\lambda^a(\sigma)\,\big(\,\partial_\nu\,{\mathlarger{\mathlarger{\mathpzc e}}}_\mu^b(\sigma)-\partial_\mu
\,{\mathlarger{\mathlarger{\mathpzc e}}}_\nu^b(\sigma)\big)
\Big)\, .\nonumber
\end{eqnarray}
In the case under consideration, the torsion is a 3-form and
the more elegant expressions \eqref{aasiais}  can be achieved by anti-symmetrizing w.r.t. the 
Greek indices and using the  formula  
\begin{eqnarray}\label{jassas}
&q_{ab}\,{\mathlarger{\mathlarger{\mathpzc e}}}^{a}_{\,[\lambda}(\sigma)\,\partial_\mu^{\vphantom a}\,{\mathlarger{\mathlarger{\mathpzc e}}}^b_{\nu]}(\sigma)-\frac{1}{2}\ \sum_{\sigma'=\pm}
F_{abc}(\sigma \sigma|\sigma')
\ {\mathlarger{\mathlarger{\mathpzc e}}}^a_{[\lambda}(\sigma)\,{\mathlarger{\mathlarger{\mathpzc e}}}^b_{\mu}(\sigma)\,{\mathlarger{\mathlarger{\mathpzc e}}}^c_{\nu]}(\sigma') =0\nonumber
\end{eqnarray}
valid for  both  choices  of $\sigma=\pm$. The later is an immediate consequence of the 
Maurer-Cartan structure
equations \eqref{shhssy}.

\bigskip
Formulae \eqref{ajssuaaus} and \eqref{aasiais}  can be made more transparent using the notation 
$\tilde{  F}_{abc}(\sigma\,\sigma'\,\sigma'')$:
\bea
F_{abc}(\sigma\,\sigma'|\sigma'')\,=\,\
\re^{\frac{\ri\pi}{4}(\sigma+\sigma'-\sigma'')}\,{\tilde F}_{abc}(\sigma\,\sigma'\,\sigma'')\ .\nonumber
\eea
The advantage of 
 ${\tilde  F}_{abc}(\sigma\,\sigma'\,\sigma'')$ compared to 
 $F_{abc}(\sigma\,\sigma'|\sigma'')$  is that  it is  a completely antisymmetric symbol  w.r.t. the pair permutations
$(a,\sigma)\leftrightarrow (b,\sigma')$  and $(b,\sigma')\leftrightarrow (c,\sigma'')$:
\bea
{\tilde  F}_{abc}(\sigma\,\sigma'\,\sigma'')=-{\tilde F}_{bac}(\sigma'\,\sigma\,\sigma'')=-{\tilde F}_{acb}(\sigma\,\sigma''\,\sigma')
\ .\nonumber
\eea
Then  \eqref{ajssuaaus},\,\eqref{aasiais} can be re-written as
\bea
G_{\mu\nu\phantom{\lambda}}&=&\frac{\ri}{4{\rm g}^2}\ \sum\limits_{\sigma=\pm}\, \sigma\, \nonumber
q_{ab}\ { {\mathpzc E}}^a_\lambda(\sigma)\,{ {\mathpzc E}}^b_\mu(\sigma) \\[0.2cm]
H_{\lambda\mu\nu}&=&\frac{1}{4{\rm g}^2}\,
\sum\limits_{\sigma,\sigma',\sigma''=\pm} {\rm sgn}(\sigma+\sigma'+\sigma'')\,{\tilde  F} _{abc}(\sigma\,\sigma'\,\sigma'')\ 
{ {\mathpzc E}}^a_\lambda(\sigma)\,{ {\mathpzc E}}^b_\mu(\sigma')\, { {\mathpzc E}}^c_\nu(\sigma'')\, ,\nonumber
\eea
where we also  use 
\bea
{{\mathpzc E}}_\mu^a(\sigma)\,\equiv \,
\re^{-\frac{\ri\pi}{4}\sigma}\ {\mathlarger{\mathlarger{\mathpzc e}}}_\mu^a(\sigma)\ .\nonumber
\eea

\subsubsection*{Ricci  tensor }

Let ${\mathsf R}_{\mu\nu}$ be the Ricci tensor built from the affine connection ${\mathsf\Gamma}$ \eqref{Connection1}.
 For practical purposes, it is useful to 
express it in terms of the symmetric Ricci tensor $R_{\mu\nu}$ associated
with the Levi-Civita connection.\footnote{Below, the
Ricci tensor is defined as $R_{\mu\nu}={R^{\lambda}}_{\mu\lambda\nu}$ where ${R^{\rho}}_{\lambda\mu\nu}$ is the
Riemann tensor 
$${R^{\lambda}}_{\mu\rho\nu}=
\partial_\rho\Gamma^{\lambda}_{\mu\nu}-\partial_\nu\Gamma^{\lambda}_{\mu\rho}+
{\Gamma^\lambda}_{\sigma\rho}{\Gamma^\sigma}_{\mu\nu}-{\Gamma^\lambda}_{\sigma\nu}{\Gamma^\sigma}_{\mu\rho}$$
and  $\Gamma^{\sigma}_{\mu\nu}=\Gamma^{\sigma}_{\nu\mu}$ stands for the  Christoffel symbols for the Levi-Civita connection.}
 Using the results from the work \cite{Valent:2009nv}  one can show that
\bea\label{iasusausauas}
{\textstyle\frac{1}{2}}\, {\mathsf R}_{(\mu\nu)}&=&R_{\mu\nu}-{\textstyle \frac{1}{4}}\, {H_\mu}^{\sigma\rho} H_{\sigma\rho\nu}={\textstyle\frac{1}{8}}\, 
\big(1-(\varepsilon_1-\varepsilon_2)^2\big)\, \big(1-(\varepsilon_1+\varepsilon_2)^2\,\big) \,
\sum_{\sigma=\pm}
q_{ab} \, { {\mathlarger{\mathlarger{\mathpzc e}}}}^{a}_\mu(\sigma) {{\mathlarger{\mathlarger{\mathpzc e}}}}^{b}_\nu(-\sigma)\nonumber \\
&  -&
\nabla_\mu W_\nu-\nabla_\nu W_\mu\\[0.2cm]
{\textstyle\frac{1}{2}}\, 
{\mathsf R}_{[\mu\nu]}&=& {\textstyle \frac{1}{2}}\ \nabla_\lambda{H^\lambda}_{\mu\nu}=
{\textstyle\frac{1}{8}}\, 
\big(1-(\varepsilon_1-\varepsilon_2)^2\big)\, \big(1-(\varepsilon_1+\varepsilon_2)^2\,\big) \,
\sum_{\sigma=\pm}
q_{ab} \,\sigma
 { {\mathlarger{\mathlarger{\mathpzc e}}}}^{a}_\mu(\sigma) {{\mathlarger{\mathlarger{\mathpzc e}}}}^{b}_\nu(-\sigma)\nonumber\\
&+&W_\lambda \,{H^\lambda}_{\mu\nu}
+\partial_\mu W_\nu-
\partial_\nu W_\mu\ . \nonumber
\eea
Here  
\bea\label{jassaty}
&W_\mu=-\frac{1}{2}\ \partial_\mu  \log\big({\det\hat{{\boldsymbol \Omega}}_\sigma}\big)+w_\mu
\eea
with ${\boldsymbol \Omega}_\sigma$ given by \eqref{defomega1} and
\bea\label{akksjsasausa}
&w_\mu=\pm \frac{\rm i}{4}\ 
{\mathlarger{\mathlarger{\mathpzc e}}}^{a}_\mu(\pm)
 \  {f_{ab}}^c\
{(  \varepsilon_1\, 
{\bar {\cal R }} -
  \varepsilon_2\,{\cal R})^b}_c\ .\nonumber
\eea
The  last  formula holds true for any choice of the sign $\pm$ and we use the notation
 \bea 
{{\bar {\cal R}}^b}{}_c=
{({\cal U}^{-1} {\cal R \,U})^{b}}_c= {({\cal U}^{-1})^b}_{b'}\, {{\cal R}^{b'}}_{c'}\, {{\cal U}^{c'}}_c\ ,\nonumber
\eea
where
 ${{\cal U}^b}_a$  stands  for   the  $ D\times D$  matrix of the group element ${\boldsymbol U}$
in the adjoint representation:
\bea
{\boldsymbol U}\,{\tt t}_a\,{\boldsymbol U}^{-1}={\tt t}_b\  {{\cal U}^b}_a\ .\nonumber
\eea

 \subsubsection*{ 1-loop renormalization of  the  Klim\v{c}\'{i}k  NLSM}

In the path-integral quantization, the general NLSM \eqref{aisiaosioa}  should be equipped with a  UV cutoff.
A consistent removal of the UV  divergences requires  that  the ``bare''   target space metric and torsion potential
be given a certain dependence on the cutoff momentum $\Lambda$. To the first  perturbative order 
in the Planck constant $\hbar$
the RG flow equations  are given by \cite{Friedan:1980jf,Fradkin:1985ys,Callan:1985ia}
\bea\label{kasauaua}
\partial_\tau G_{\mu\nu}&=&-\hbar\,
\Big(R_{\mu\nu}-\frac{1}{4}\ {H_\mu}^{\sigma\rho} H_{\sigma\rho\nu}+\nabla_\mu V_\nu+\nabla_\nu V_\mu\Big)
+O(\hbar^2)\nonumber\\[-0.2cm]
\\[-0.2cm]
\partial_\tau B_{\mu\nu}&=&-\hbar\,
\Big( \,-\frac{1}{2}\ \nabla_\lambda{H^\lambda}_{\mu\nu}+V_\lambda \,{H^\lambda}_{\mu\nu}+\partial_\mu
 \Lambda_\nu-
\partial_\nu \Lambda_\mu\Big)+O(\hbar^2)\ ,\nonumber
\eea
where  
$
\partial_\tau\equiv 2\pi\,  {\Lambda}\,{\textstyle \frac{{\textstyle \partial}}{{\textstyle \partial \Lambda}}}\,.
$ The infinitesimal variation of the Klim\v{c}\'{i}k metric and torsion potential, 
assuming
that the combinations of the couplings $\frac{\varepsilon_2}{\varepsilon_1}$, ${\rm g}^2\varepsilon_1$
are kept fixed, can be expressed as
\bea
\delta  G_{\mu\nu}&=&+
\frac{\delta \varepsilon_1}{   4{\rm g}^2\varepsilon_1}\
\  \sum_{\sigma=\pm}
q_{ab} \, { {\mathlarger{\mathlarger{\mathpzc e}}}}^{a}_\mu(\sigma) 
{{\mathlarger{\mathlarger{\mathpzc e}}}}^{b}_\nu(-\sigma) \nonumber\\
\delta  B_{\mu\nu}&=&
-\frac{\delta \varepsilon_1}{   4{\rm g}^2\varepsilon_1}\
\ \sum_{\sigma=\pm}
q_{ab} \, \sigma\, { {\mathlarger{\mathlarger{\mathpzc e}}}}^{a}_\mu(\sigma) {{\mathlarger{\mathlarger{\mathpzc e}}}}^{b}_\nu(-\sigma)\ .\nonumber
\eea
With the  explicit formulae for the Ricci tensor \eqref{iasusausauas}, it is easy to see  that
 the general RG flow equations \eqref{kasauaua} are satisfied if 
 $V_\mu=\Lambda_\mu=W_\mu$  with $W_\mu$  given by \eqref{jassaty}.
   Also it follows that
  the evolution of the bare couplings  under a change in $\Lambda$
 is described by the system of ordinary differential  equations \eqref{hassasaty}.

\section{Appendix\label{AppC}}

In this Appendix we provide the explicit relation between the flat connection \eqref{siqjkhdsjh2} for 
the case of the Fateev model ($\mathfrak{G}=SU(2)$)
and that given in the work \cite{Lukyanov:2012zt}. 
\bigskip

In that work a more general four parameter deformation of the $SU(2)$ principal chiral field is considered which contains
the Fateev model as a two-parameter subfamily. The deformation parameters were denoted by
$(\eta,\nu^{(L)},\sigma,q)$ and,  for the case of the Fateev model, $\nu^{(L)}$ together
with $\sigma$ should be set to zero:
\be
\nu^{(L)}=\sigma=0\, .\nonumber
\ee
Here the superscript $L$ has been used to distinguish the parameter $\nu$ in
ref.\cite{Lukyanov:2012zt} with the one from this work.
The remaining two parameters $\eta$ and $q$ are related to 
 $\kappa$ and $\nu$ in \eqref{kappdef} as
\be
\kappa=\frac{\vartheta^2_2(0,q^2)}{\vartheta^2_3(0,q^2)}\,, \qquad\qquad 
 \nu=- \ri\  \frac{\vartheta_1(\ri\eta,q^2)}{\vartheta_4(\ri\eta,q^2)}\,,\nonumber
\ee
where $\vartheta_a$ stand for the conventional theta functions. 
In ref.\cite{Lukyanov:2012zt} the same co-ordinates $v$ and $w$
that appear in the Euler decomposition \eqref{skdjfsd} are used, while $\phi$ from
 \eqref{oqiwoeiw} is replaced by $u$, such that
 \bea
\tanh(\phi)=\frac{\vartheta_2(u,q^2)\vartheta_3(0,q^2)}{\vartheta_3(u,q^2)\vartheta_2(0,q^2)}\ \ \ \ \ \ \ \ \ \ (0<u<\pi)\ .\nonumber
\eea

The flat connection ${\boldsymbol A}^{(L)}_\pm$  found in \cite{Lukyanov:2012zt}
is defined by eqs.\,(1.6),\,(2.7) and (2.10)-(2.14) from that work, where
$\lambda$ is the spectral parameter and, for the Fateev model, $\eta_+=\eta_-=\eta$
and $\phi_\pm=0$.
Formulae (2.7),\,(2.10) 
involve the vielbein $e^a_\mu$ $(\mu=u,v,w)$, which in turn are given by eqs.\,(2.28)-(2.32).
Here, for the convenience of the reader, we reproduce the main equations
needed for the computation of ${\boldsymbol A}^{(L)}_\pm$ 
specialized to the Fateev model.

The non-vanishing components of the vielbein are given by
\bea
e^3_u &=&\phantom{\pm\,} \frac{\ri}{{\rm g}}\, \frac{\vartheta_2(\ri\eta,q)\,\vartheta_1'(0,q)}{\vartheta_1(\ri\eta,q)\,\vartheta_2(0,q)}\nonumber \\[0.2cm]
e^\pm_v &=& \mp\, \frac{\ri}{{\rm g}}\,\frac{\vartheta_4(0,q^2)\,\vartheta_4(\ri\eta\pm u,q^2)}{\vartheta_4(u,q^2)\,\vartheta_4(\ri\eta,q^2)} 
\nonumber\\[0.2cm]
e^\pm_w &=& \pm\, \frac{\ri}{{\rm g}}\,\frac{\vartheta_4(0,q^2)\,\vartheta_1(\ri\eta\pm u,q^2)}{\vartheta_4(u,q^2)\, \vartheta_1(\ri\eta,q^2)}\ .
\nonumber
\eea
Note that, with these  expressions at hand, it is simple to  re-write the Lagrangian of the Fateev model  
in terms of the parameters $(\eta, q)$ and the co-ordinates $X^\mu=(u,v,w)$ since
\be
{\cal L}_F=2\,G_{\mu\nu}\,\partial_+ X^\mu\partial_-X^\nu\nonumber
\ee
and the non-zero components of the metric tensor $G_{\mu\nu}$ are
\be
G_{uu} = (e^3_u)^2\,, \qquad G_{vv}=e^+_ve^-_v\,, \qquad G_{ww}=e^+_w e^-_w\,,\qquad G_{vw}=\tfrac{1}{2}\,(e^+_v\,e^-_w+e^-_v\,e^+_w)\,.\nonumber
\ee
The connection is constructed from the
 matrix valued 1-form ${\bm \zeta}_\mu(\lambda)$
defined by
\be
{\bm \zeta}_\mu(\lambda)=f_3(\lambda)\,e^3_\mu\,\sigma^3 + 
f_+(\lambda)\,e^+_\mu\,\sigma^- +  f_-(\lambda)\,e^-_\mu\,\sigma^+\,,\nonumber
\ee
where $\sigma^3$ and $\sigma^\pm=\tfrac{1}{2}(\sigma^1\pm\ri\sigma^2)$ are the standard Pauli matrices, while
\bea
f_+(\lambda)&=&-f_-(-\lambda)=-\frac{{\rm g}}{2}\ 
\frac{\vartheta_1(u-\frac{\lambda}{2},q)\,\vartheta_1(\ri\eta,q)\,\vartheta_2(0,q)}{\vartheta_1(u,q)\,\vartheta_2(\ri\eta,q)
\,\vartheta_1(\frac{\lambda}{2},q)} \nonumber \\[0.2cm]
f_3(\lambda)&=&-\frac{{\rm g}}{2}\ \frac{\vartheta_1(\ri\eta,q)\,\vartheta_2(0,q)\,\vartheta_1'(\frac{\lambda}{2},q)}{\vartheta_2(\ri\eta,q)\,
\vartheta_1'(0,q)\,\vartheta_1(\frac{\lambda}{2},q)}\ \,.\nonumber
\eea
In terms of this 1-form, 
the connection components ${\boldsymbol A}^{(L)}_\pm$ are expressed as
\bea
{\boldsymbol A}^{(L)}_+&=&\frac{1}{2\ri}\,\sum_{\mu}\,\big(\bm{\zeta}_\mu(\ri\eta+\lambda)+\sigma^2\,
\bm{\zeta}_\mu(\ri\eta-\lambda)\,\sigma^2\,\big)\,\partial_+ X^\mu \nonumber\\[0.2cm]
{\boldsymbol A}^{(L)}_-&=&\frac{1}{2\ri}\,\sum_{\mu}\,\big(\bm{\zeta}_\mu(\ri\eta+\lambda-\pi)+\sigma^2\,
\bm{\zeta}_\mu(\ri\eta-\lambda+\pi)\,\sigma^2\,\big)\,\partial_- X^\mu \,,\nonumber
\eea
where $X^\mu=(u,v,w)$.
One should keep in mind that the zero curvature representation
in \cite{Lukyanov:2012zt} is
\be
\big[\partial_++{\boldsymbol A}_+^{(L)},\,\partial_-+{\boldsymbol A}_-^{(L)}\,\big]=0\,,\nonumber
\ee
which differs from the convention used in this work \eqref{ZCR1} by the overall sign of $\bm{A}_\pm$.

The gauge transformation that maps 
the flat connection ${\boldsymbol A}^{(L)}_\pm$ 
 to the one in \eqref{siqjkhdsjh2},\,\eqref{currentsK} with $\bm{U}$ understood as a matrix in the 
 fundamental representation of $SU(2)$ (i.e., ${\tt h}=\sigma^3$, ${\tt e}_\pm=\sigma^\pm$),
 is described as follows:
\begin{equation}
\partial_\pm-{\boldsymbol A}_\pm\,=\,{\boldsymbol S}\,\Big(\partial_\pm + {\boldsymbol A}^{(L)}_\pm\,\Big)\,{\boldsymbol S}^{-1}\,,
\nonumber
\end{equation}
where 
\bea
{\boldsymbol S}\,=\,\sqrt{\frac{\vartheta_4(\lambda,q^2)\vartheta_4(0,q^2)}{2\vartheta_1(\lambda,q^2)\vartheta_4(u,q^2)}}\,\left(\begin{array}{cc}
\re^{\frac{\ri w}{2}}\ \frac{\vartheta_2(\,\frac{1}{2}(\lambda-u),\,q)}{\vartheta_3(\frac{\lambda}{2},q)}   &
  \ri\, \re^{\frac{\ri w}{2}}\ \frac{\vartheta_2(\,\frac{1}{2}(\lambda+u),\,q)}{\vartheta_3(\frac{\lambda}{2},q)}  \\[0.5cm]
\ri\,\re^{-\frac{\ri w}{2}}\ \frac{\vartheta_1(\,\frac{1}{2}(\lambda-u),\,q)}{\vartheta_4(\frac{\lambda}{2},q)} &  
\re^{-\frac{\ri w}{2}}\ \frac{\vartheta_1(\,\frac{1}{2}(\lambda+u),\,q)}{\vartheta_4(\frac{\lambda}{2},q)} \end{array}\right)
\nonumber
\eea
and ${\boldsymbol S}^{-1}=\sigma_2\,{\boldsymbol S}^T\sigma_2\ (\det{ \boldsymbol S}=1)$. \!The parameters $\rho_\pm$  are 
expressed in terms of the spectral parameter $\lambda$ as
\bea
\frac{\rho_+}{\rho_-}=
\frac{\vartheta_3^2(\frac{\lambda}{2},q) }{\vartheta_4^2(\frac{\lambda}{2},q)}\ ,
\ \ \ \ \ \ \ \ \ \  \ 
\rho_+\rho_-=\frac{\vartheta^2_4(\frac{\ri\eta}{2},q) }{\vartheta^2_3(\frac{\ri\eta}{2},q)}\ .
 \nonumber
\eea
Finally note that the original deformation parameters $\varepsilon_1$, $\varepsilon_2$ 
in the Lagrangian \eqref{oaspsaop}
are related to $q$ and $\eta$ as
\be
\varepsilon_1=\frac{\vartheta_4^2(\ri\eta,q^2)\,\vartheta_3(0,q^2)\,\vartheta_2(0,q^2)}{
                                       \vartheta_4^2(0,q^2)\,\vartheta_3(\ri\eta,q^2)\,\vartheta_2(\ri\eta,q^2)}\,, \qquad
\varepsilon_2=-\frac{\vartheta_1^2(\ri\eta,q^2)\,\vartheta_3(0,q^2)\,\vartheta_2(0,q^2)}{
                                       \vartheta_4^2(0,q^2)\,\vartheta_3(\ri\eta,q^2)\,\vartheta_2(\ri\eta,q^2)}\, .\nonumber
\ee

\end{document}